\documentclass[11pt,a4paper]{article}
\usepackage{jheppub}
\usepackage{amsmath}
\usepackage{amssymb}
\usepackage{tikz}
\usepackage{pgfplots}
\usepackage{graphicx}

\newcommand{\x}{\mathbf{x}}
\newcommand{\p}{\mathbf{p}}
\def\lf{\left\lfloor}   
\def\rf{\right\rfloor}

\newcommand{\bra}[1]{\langle #1|}
\newcommand{\ket}[1]{|#1\rangle}

\newcommand{\be}{\begin{equation}}
\newcommand{\ee}{\end{equation}}
\newcommand{\bea}{\begin{eqnarray}}
\newcommand{\eea}{\end{eqnarray}}
\newcommand{\reef}[1]{(\ref{#1})}
\newcommand{\mt}[1]{\textrm{\tiny #1}}
\newcommand{\non}{\nonumber \\}
\newcommand{\eg}{{\it e.g.,}\ }
\newcommand{\ie}{{\it i.e.,}\ }

\def\({\left(} \def\){\right)}
\def\[{\left[} \def\]{\right]}
\def\al{\alpha} \def\bt{\beta}

\title{Dynamics of the Area Law of Entanglement Entropy}
\author{Stefan Leichenauer,}
\author{Mudassir Moosa,}
\author{and Michael Smolkin}

\affiliation{Center for Theoretical Physics and Department of Physics,\\
University of California, Berkeley, CA 94720, U.S.A.} 
\affiliation{Lawrence Berkeley National Laboratory, Berkeley, CA 94720, U.S.A.} 

\emailAdd{sleichen@berkeley.edu}
\emailAdd{mudassir.moosa@berkeley.edu}
\emailAdd{smolkinm@berkeley.edu}

\abstract{We study the evolution of the universal area law of entanglement entropy when the Hamiltonian of the system undergoes a time dependent perturbation. In particular, we derive a general formula for the time dependent first order correction to the area law under the assumption that the field theory resides in a vacuum state when a small time-dependent perturbation of a relevant coupling constant is turned on. Using this formula, we carry out explicit calculations in free field theories deformed by a time dependent mass, whereas for a generic QFT we show that the time dependent first order correction is governed by the spectral function defining the two-point correlation function of the trace of the energy-momentum tensor. We also carry out holographic calculations based on the HRT proposal and find qualitative and, in certain cases, quantitative agreement with the field theory calculations.

}

\begin{document}
	\maketitle
	
	\section{Introduction}
	
In recent years, the study of the time-dependence of entanglement entropy has received a considerable attention \cite{CC-particle,Balasubramania_one, Balasubramanian_two, Maldacena-Hartman,HL-one, HL-two,   tsunami, Asplund, CLM, Rangamani} . In the simplest scenario one studies the evolution of the spatial entanglement entropy under the assumption that the system begins in the vacuum state in the asymptotic past while a relevant (i.e., positive mass dimension) coupling constant in the Hamiltonian is given a non-trivial time dependence which asymptotes to a constant value in the far future. 
	
Although the assumption that the system starts in the vacuum state might seem to be an oversimplification, certain questions about the entanglement may still be addressed without loss of generality. For example, in quantum field theory the entropy of a region is always UV divergent \cite{Bomb86,Sred93}, whereas the structure of these divergences is state independent, see \eg \cite{misha_myers_hung}. Hence, assuming that the system resides in the vacuum state is enough to determine the pattern of divergences in $any$ state. Indeed, this pattern is fixed by the entanglement of degrees of freedom near the UV cutoff scale, $\delta$, and therefore the characteristic scale of an excited state plays no role in this regime. Furthermore, the dynamics of the divergent part of the entanglement entropy in a time dependent setup is not trivial, since the asymptotic vacua, in general, are not identical.
	
Even though the time dependence of the full entanglement entropy in a generic setup is very complicated, we expect that evolution of the UV divergences is quasi-adiabatic, and therefore tractable.  Our expectation rests on the locality of UV divergences when all length scales, $\ell_i$, including the characteristic time scale of the change in couplings, are much longer than the UV cut off, $\ell_i\gg \delta$, \ie when the system is found in the physical regime.\footnote{In principle, one could study situations when the coupling  varies over time scales of order the UV cut off. However, the interpretation of these results might be questionable, and we leave this regime outside the scope of the current work.} In particular, the divergent part of the entropy at time $t$ will only depend on the instantaneous value of the coupling and its time derivatives.
	
There is an additional outcome that stems directly from the locality of divergences. It is now a well-established fact that entanglement entropy is sensitive to the geometry of the background and shape of the entangling surface \cite{solo,Hung:2011xb,Fur13,Dong:2013qoa,Camps:2013zua,Allais:2014ata,Mezei:2014zla,Faulkner:2015csl,Bianchi:2015liz}.  The full answer can be expressed as a sum of certain geometric structures with coefficients fixed by QFT data. Not all of these structures are necessarily local, however those which are multiplied by divergent coefficients as $\delta\to 0$ should be local. 
	
The ubiquitous `area law' \cite{Bomb86,Sred93} is probably the most prominent and simple example of a local geometric structure. In a static $d$-dimensional CFT it is proportional to $\delta^{2-d}\mathcal{A}$  since $\delta$ in this case is the only admissible scale at hand, while entropy is dimensionless and area of the surface, $\mathcal{A}$, has length dimension $d-2$. Therefore the `area law' term is not universal in a CFT -- it is sensitive to the choice of regularization scheme. However, the presence of an additional scale in a non-conformal field theory may overturn this conclusion and result in a {\it universal} `area law' of entanglement, \eg a logarithmically divergent term proportional to the area multiplied by an appropriate power of some built-in scale. 

In this work we focus on the evolution of the `area law' terms when a QFT undergoes a quantum quench, and calculate the time-dependence of the universal coefficients in a variety of examples. To make the field theory analysis analytically tractable, we expand the entanglement entropy to linear order in the time dependent coupling $\lambda(t)$.
This is not necessary to recover the exact pattern of the universal divergences in the holographic approach, since it is completely fixed by the asymptotic region of the dual geometry.  Even though our focus is on the `area law' only, we expect qualitatively similar behavior from the subleading divergences as well.

If the mass dimension of $\lambda(t)$ is $\alpha$, then for $\alpha = (d-2)/k$ we may have a universal term proportional to $\lambda^k(t)\mathcal{A}$. Similarly, if $\alpha = (d-2-l)/k $ with $l$ being a non-negative integer, then a term proportional to $\lambda^{k-m}(t) \partial_t^l \lambda^m(t)\mathcal{A}$  with $1\leq m\leq k$ is allowed. It may happen that for a given $\alpha$ there is more than one choice of $l$ and $k$, in which case there is a proliferation of possible universal structures. Generically, one would expect all allowed terms to appear, and with independent coefficients. Note, however, that for a given $\alpha$ there are only finitely many allowed terms.

In sections \ref{sec-holo} and \ref{sec-freefield} we carry out explicit calculations for a set of special values of $k$ and $l$ in holographic and free field theories. In both cases we observe the quasi-adiabatic behavior that we had argued for. Moreover, in some cases our holographic and free field calculations exactly match. Such an agreement between the time-dependent terms of entanglement entropy in the weakly and strongly coupled regimes resembles similar match in the static case \cite{RS4,Faulkner:2014jva} and can be attributed to the universality of certain two- and three-point correlation functions in a CFT.  In particular, we observe explicit match when $\alpha=1,2$ and $k=2$ in various dimensions.\footnote{In fact, there is also a trivial match when $k=1$. The universal term in this case is independent of $\lambda_0$, and therefore the final answer remains intact if we set $\lambda_0=0$. However, if the theory is conformal at $\lambda_0=0$ such terms must vanish independently of the values of $\alpha$ and $l$ for the same reason as in the static case \cite{RS1}, see section \ref{sec-freefield}.} These values of $\al$ correspond to the Dirac and scalar mass operators, respectively. Recalling now that $\alpha = (d-2-l)/k$ results in the following cases explicitly elaborated in the text: 
\begin{itemize}
\item $\alpha =1$, $d=4$, $l=0$. This case corresponds to a 4D massive Dirac field. Up to a numerical factor the universal entanglement entropy is given by $\sim  \lambda^2(t)\mathcal{A}$, where $\lambda(t)$ is the time-dependent mass of the Dirac field.
\item $\alpha =2$, $d=6$, $l=0$. This case of a 6D massive scalar field is analogous to the previous case, \ie the entanglement entropy is given by $\sim  \lambda^2(t)\mathcal{A}$, except that $\lambda(t)$ is mass squared this time.
\item $\alpha = 1$, $d=6$, $l=2$. In six dimensions, there is a universal term involving two time derivatives of the Dirac mass $\sim \lambda(t) \ddot\lambda(t)\mathcal{A}$. We calculate the numerical coefficient for the first time both holographically and in the free field theory, and find exact agreement.
\item $\alpha =2$, $d=8$, $l=2$. The behavior of the universal term for the massive scalar field in eight dimensions $\sim \lambda(t) \ddot\lambda(t)\mathcal{A}$ resembles the massive Dirac field in six dimensions, and again we calculate the precise numerical coefficient holographically and in free field theory, finding agreement.
\end{itemize}

The remainder of this paper is organized as follows. In Section~\ref{sec-holo} we build on the holographic proposal \cite{RT,HRT} to compute the time-dependent universal entanglement entropy for holographic field theories. Since we are only interested in divergent terms, the gravity calculation is localized in the asymptotic region of the dual geometry; this feature allows us to calculate the universal coefficients exactly even though the geometry is only determined perturbatively. We calculate the coefficients of the terms for which agreement with the free fields is expected, as well as for the terms where agreement is not anticipated. In the latter case the answer on the field theory side depends on more than just the universal correlation functions of the underlying CFT. 

In Section~\ref{sec-freefield} we derive a general expression for the time-dependent first order correction to the `area law', and show that for a generic QFT it can be expressed in terms of the spectral function defining the two-point correlation function of the trace of the energy-momentum tensor.  We use these formulas to calculate the universal `area law' for massive free fields with time dependent masses.  We end with conclusions in Section~\ref{sec-discuss}, and a number of Appendices detailing some of the more technical aspects of the calculation.


	\section{Holographic Calculation}\label{sec-holo}
	
	\subsection{Setup}
	
	We consider a CFT in $d$ spacetime dimensions on the boundary, dual to Einstein gravity in the $d+1$-dimensional bulk. The computation of the entanglement entropy of some boundary region, $B$, can be done using HRT prescription \cite{RT, HRT}. The prescription tells us to find the codimension-$2$ surface in the bulk with extremal area, $\Sigma_{B}$, subject to the condition that $B$ and $\Sigma_B$ are homologous. If there are multiple extremal surfaces, we must choose the one with the minimum area. We will call this minimal extremal surface the HRT surface from now on. The entanglement entropy of $B$ and the area of the HRT surface are related by
	\begin{equation}
	S = \frac{\text{Area}(\Sigma_{B})}{4G_{d+1}}. \label{hrt}
	\end{equation}
	
	Beginning with the vacuum state of the CFT, we turn on a relevant operator $\mathcal{O}$ of dimension $\Delta <d$ at $t=0$ by giving it a nonzero, time-dependent, spatially homogenous coupling $\lambda(t)$. It will be useful to express our results in terms of the mass dimension of the coupling, $\alpha \equiv d-\Delta$. The operator $\mathcal{O}$ is dual to a scalar field $\Phi$ in the bulk of mass $m^2 = \Delta(\Delta-d)$ \cite{adscft_one, adscft_two}. A nonzero coupling $\lambda(t)$ corresponds to nontrivial boundary conditions for $\Phi$. Through the Einstein-scalar field equations, this will change the bulk metric and therefore affect the calculation of the HRT surface and its area.
	
	The Einstein-scalar equations are given by the variation of the bulk action \cite{misha_myers_hung}
	\begin{align}
	I(g_{\lambda\nu},\Phi) =& \frac{1}{16\pi G_{d+1}} \int d^{d+1}x\mbox{ }\sqrt{-g}\mbox{ }\Bigg\{R - \frac{1}{2}g^{\lambda\nu}\partial_{\lambda}\Phi\partial_{\nu}\Phi - V(\Phi) \Bigg\} \label{bulk_action}\\
	V(\Phi) =& - d(d-1) + \frac{1}{2}m^{2}\Phi^{2} + \frac{\kappa}{6}\Phi^{3} +\frac{\omega}{24}\Phi^{4} + O(\Phi^{5})
	\end{align}
	where we have set $L_{AdS} = 1$. We have chosen to normalize the bulk scalar action so that there is an overall factor of $1/16\pi G_{d+1}$ in front; this amounts to a normalization condition on the operator $\mathcal{O}$. In particular, with this normalization, we can read off Newton's constant from the two-point function of $\mathcal{O}$ \cite{Klebanov_Witten}:
	\begin{equation}
	\langle \mathcal{O}(x)\mathcal{O}(y) \rangle =\mbox{ }\frac{2\Delta - d}{8\pi G_{d+1}}\mbox{ }\frac{\Gamma(\Delta)}{\Gamma(\Delta - d/2)}\mbox{ }\frac{1}{\pi^{d/2}|x-y|^{2\Delta}} \label{def_G}.
	\end{equation}
This will allow us to make contact with the free field calculations in Section~\ref{sec-freefield}.\footnote{Alternatively, we could have left the normalization unspecified at the cost of introducing another parameter $\eta$ which multiplies the bulk scalar field action. The normalization we have chosen is convenient because it simplifies the Einstein-scalar equations of motion.} The bulk interaction parameters $\kappa$, $\omega$, etc., are similarly determined by the three, four, and higher-point functions of $\mathcal{O}$. The most interesting terms in the entanglement entropy are the ones which depend only on the two-point function, since these may be the same across theories once the normalization has been fixed using \eqref{def_G}. Thus we should pay special attention below to those terms which depend only on $G_{d+1}$.

	 The coupled Einstein-scalar equations are
	\begin{align}
	R_{\mu\nu} = \frac{1}{2}\partial_{\mu}\Phi\partial_{\nu}\Phi + \frac{1}{d-1}g_{\mu\nu}V&(\Phi)  \label{einstein_equation}\\
	\frac{1}{\sqrt{-g}}\partial_{\mu}\Big(\sqrt{-g}g^{\mu\nu}\partial_{\nu}\Phi\Big) - \frac{\delta}{\delta\phi} V(\Phi) =& 0 \label{scalar_equation}
	\end{align}
	Following \cite{Witten_Graham}, we choose our bulk `radial' coordinate $z$ to be orthogonal to all of the boundary coordinates. This corresponds to the gauge fixing
	\begin{equation}
	g_{zt} = g_{zi} = 0,
	\end{equation}
	where $x^{i}$ for $i=1,...,d-1$ are the spatial coordinates of the boundary. We also rescale the bulk coordinate to fix $g_{zz} = \frac{1}{z^{2}}$.
	
	Since our quench is homogenous and isotropic, the bulk geometry must be homogenous and isotropic as well. Then we can write the bulk metric as
	\begin{equation}
	ds^{2} = \frac{1}{z^{2}} \Big( dz^{2} - f(t,z)dt^{2} + h(t,z) \sum_{i=1}^{d-1}dx_{i}^{2}\Big) \label{bulk_metric}
	\end{equation}
	with the boundary conditions that $f(t,z=0) = h(t,z=0) = 1$, and $f(t<0,z) = h(t<0,z) = 1$.
	
	With this choice of metric, \eqref{einstein_equation} and \eqref{scalar_equation} take the following form\footnote{These equations are the $zt$-component \eqref{equation_two} and $tt$-component \eqref{equation_three} of the Einstein equation, and the scalar field equation of motion \eqref{equation_four}. We will not need the remaining equations in our analysis.}:
	\begin{align}
	0&=\frac{d-1}{4}\frac{\dot{h}}{h}\frac{h'}{h} + \frac{d-1}{4}\frac{\dot{h}}{h}\frac{f'}{f} - \frac{d-1}{2}\frac{\dot{h}'}{h} - \frac{1}{2}\dot{\Phi}\Phi',  \label{equation_two}\\
	0&= \frac{d-1}{4}\Big(\frac{\dot{h}}{h}\Big)^{2} -\frac{d}{2}\frac{f'}{z} + \frac{d-1}{4}\frac{\dot{f}}{f}\frac{\dot{h}}{h} - \frac{1}{4}\frac{f'^{2}}{f} + \frac{d}{z^{2}}f - \frac{d-1}{2}\frac{f}{z}\frac{h'}{h}\nonumber\\
	&~~~~~ - \frac{d-1}{2}\frac{\ddot{h}}{h} + \frac{d-1}{4}\frac{f'h'}{h} + \frac{1}{2}f'' - \frac{1}{2}\dot{\Phi}^{2} + \frac{1}{d-1}\frac{f}{z^{2}}V(\Phi),  \label{equation_three}\\
	0&=z^{2}\Phi'' - \frac{z^{2}}{f}\ddot{\Phi} - (d-1)z\Phi' + \frac{z^{2}}{2}\dot{\Phi}\frac{\dot{f}}{f} + \frac{z^{2}}{2}\Phi'\frac{f'}{f} + (d-1)\frac{z^{2}}{2}\Phi'\frac{h'}{h} \nonumber\\
	&~~~~~- (d-1)\frac{z^{2}}{2}\frac{\dot{\Phi}}{f}\frac{\dot{h}}{h} - \frac{\delta}{\delta\phi} V(\Phi), \label{equation_four}
	\end{align}	
	The UV divergences in the entanglement entropy arise from the divergences of the area functional of the HRT surface, $\Sigma_{B}$, near the asymptotic boundary ($z=0$). To compute these divergences, we need to find the area functional, or the induced metric on $\Sigma_{B}$, near $z=0$. In \cite{misha_myers_hung}, the authors studied the solutions of \eqref{equation_two}-\eqref{equation_four} as a power series in $z$. They have shown that the solutions can be written as the sum of two separate power series. One of them is independent of the state of the boundary theory, whereas the other series solution carries information about the state of the boundary theory. They also showed that the latter solution does not contribute to the divergences in the entanglement entropy, which implies that the divergences in the entanglement entropy are state-independent (as expected).
	
	Having learned that the state-dependent power series solution does not contribute to the divergences, we will only be writing the state-independent solution in the following analysis.\footnote{This is not to say that the state-dependent terms are unimportant for the non-divergent parts of the entropy or that they are actually equal to zero in the solutions we consider, but only that we do not need to keep track of them in our analysis.} In \cite{misha_myers_hung}, all the powers of $z$ that arise in the asymptotic solution have been identified. Using their result, we make the following ansatz for the metric and the scalar field:
	\begin{align}
	h(z,t) =&\mbox{ }1 + \sum_{m=2}\sum_{n=0}\mbox{ }z^{m\alpha + n}\mbox{ }h_{m,n}(t) \label{ans_h}\\
	f(z,t) =&\mbox{ }1 + \sum_{m=2}\sum_{n=0}\mbox{ }z^{m\alpha + n}\mbox{ }f_{m,n}(t) \label{ans_f}\\
	\Phi(z,t) =&\mbox{ }z^{\alpha}\mbox{ }\sum_{m=0}\sum_{n=0}\mbox{ }z^{m\alpha + n}\mbox{ }\phi_{m,n}(t) \label{ans_p}
	\end{align}
	where $\phi_{0,0}(t)$ is equal to the time dependent coupling, $\lambda(t)$. That is,
	\begin{equation}
	\phi_{0,0}(t) = \lambda(t). \label{dic_coupling}
	\end{equation}

	\subsection{HRT Surface} \label{HRT_surface}
	
	In this work we will only consider terms which are independent of the curvature of the entangling surface. Then it will be enough to consider the entanglements of a half-space in the field theory, which means the entangling surface is a flat plane. That is, the region $B$ in the boundary is given by
	\begin{equation}
	B:\mbox{ }\mbox{ }\mbox{ }\mbox{ }x_{1}>0,\mbox{ }\mbox{ }\mbox{ }\mbox{ }\mbox{ }\mbox{ }\mbox{ }\mbox{ }t=t_{0}.
	\end{equation}
	The natural coordinates on the codimension-$2$ HRT surface in the bulk are $\{z, x^{i}\}$ for $i=2,...,d-1$, with $t$ and $x_1$ left as functions determining the position of the surface in the bulk. The HRT surface must be invariant under translation in $x^{i}$ direction, for $i=2,...,d-1$. This means $t=t(z)$ and $x_{1}=x_{1}(z)$ are functions of $z$ alone. The state is also translation-invariant in the $x_1$ direction, and so we can fix $x_{1}(z)=0$. To summarize, the HRT surface, $\Sigma_{B}$, is 
	\begin{align}
	\Sigma_{B}:\mbox{ }\mbox{ }\mbox{ }\mbox{ }x_{1}=x_{1}(z)=0,\mbox{ }\mbox{ }\mbox{ }\mbox{ }\mbox{ }\mbox{ }\mbox{ }\mbox{ }\mbox{ }t=t(z),
	\end{align}
	with the boundary condition, $t(z=0) = t_{0}$, the time at which we are computing the entropy.
	
	The area functional of the HRT surface is given by
	\begin{equation}
	\text{Area}(\Sigma_{B}) =\mbox{ }2\mathcal{A}\mbox{ }\int_{\delta}^{\infty}dz\mbox{ }\frac{h^{\frac{d-2}{2}}(z,t(z))}{z^{d-1}}\mbox{ }\sqrt{1 - f(z,t(z))(t'(z))^{2} }, \label{area_functional}
	\end{equation}
	where $\mathcal{A}$ is the area of the entangling surface in the boundary, and prime denotes the derivative with respect to $z$. Also note that we have introduced a cut-off surface near the boundary at $z=\delta$.
	
	Applying the variational principle to the above functional gives us an equation for $t(z)$. As mentioned above, we are only interested in the asymptotic behavior of the solution near $z=0$. As we will show in Appendix~\ref{app-HRT}, the lowest nontrivial power of $z$ that appears in the solution of $t(z)$ is $z^{2+2\alpha}$. That is, at the lowest order, the solution of the stationary surface is
	\begin{equation}
	t(z) =\mbox{ }t_{0} + z^{2+2\alpha}\mbox{ }t_{2+2\alpha} + \ldots, \label{asym_geodesic}
	\end{equation}
	where $t_{2+2\alpha}$ is a constant.\\
	
	By expanding the integrand of \eqref{area_functional} in $z$, we see that we can see that $t_{2+2\alpha}$ in \eqref{asym_geodesic} enters as the coefficient of $z^{4\alpha + 3-d}$. This will not affect the divergent terms in the entropy as long as
	\begin{align}
	\alpha > \frac{d-4}{4}.
	\end{align}
In the following, we will perform explicit calculations for $\alpha = 1,2$ in $d=4,5,6$, and for $\alpha=2$ in $d=8$. These values of $\alpha$ corresponds to the mass term of a Dirac fermion and scalar boson, respectively. For these values of $\alpha$ and $d$, the above inequality is true. This allows us to forget about the details of the HRT surface, and simply set $t(z) = t_{0}$ in \eqref{area_functional}. With this simplification, \eqref{area_functional} becomes
	\begin{equation}
	\text{Area}(\Sigma_{B}) =\mbox{ }2\mathcal{A}\mbox{ }\int_{\delta}^{\infty}dz\mbox{ }\frac{h^{\frac{d-2}{2}}(z,t_{0})}{z^{d-1}}. \label{area_functional_two}
	\end{equation}
The entanglement entropy of the boundary region $B$ is then given by
	\begin{equation}
	S(t_{0}) = \frac{\mathcal{A}}{2G_{d+1}}\mbox{ }\int_{\delta}^{\infty}dz\mbox{ }\frac{h^{\frac{d-2}{2}}(z,t_{0})}{z^{d-1}}, \label{hrt_two}
	\end{equation}
where we can give $G_{d+1}$ a field-theory definition by relating it to the two-point function of the operator $\mathcal{O}$ as in \eqref{def_G}, reproduced here:
	\begin{equation}
	\langle \mathcal{O}(x)\mathcal{O}(y) \rangle =\mbox{ }\frac{2\Delta - d}{8\pi G_{d+1}}\mbox{ }\frac{\Gamma(\Delta)}{\Gamma(\Delta - d/2)}\mbox{ }\frac{1}{\pi^{d/2}|x-y|^{2\Delta}}.
	\end{equation}

	\subsection{Scalar mass operator: $\alpha = 2$} \label{holographic_scalar}
	
	In this subsection, we will specialize to the case where $\alpha = 2$. In the free field case, this corresponds to a time-dependent mass:
	\begin{align}
	\mathcal{O}(x) =&\mbox{ } \phi^{2}(x) \\
	\lambda(t) =&\mbox{ }\frac{1}{2}m^{2}(t)
	\end{align}

	The entanglement entropy is given by \eqref{hrt_two}. The ansatz for $h(z,t_{0})$ given in \eqref{ans_h} becomes
	\begin{equation}
	h(z,t) =\mbox{ }1 + z^{4}h_{2,0}(t) + O(z^{6}) \label{h_scalar}
	\end{equation}
	By counting powers of $z$ in the integral \eqref{hrt_two}, we deduce that there is no time-dependent universal log-divergence in the entanglement entropy for $d=4$ and $d=5$. This result is consistent with the time-independent holographic calculations \cite{misha_myers_hung}. Now we turn to $d=6$ and $d=8$. To facilitate comparison to the free field case (wherever applicable), we will use \eqref{def_G} and the result \cite{RS4}
	\begin{equation}
	\langle \phi^{2}(x)\phi^{2}(y) \rangle_{\text{boundary}} =\mbox{ }\frac{2}{(d-2)^{2}\Omega_{d}^{2}}\mbox{ }\frac{1}{|x-y|^{2(d-2)}}
	\end{equation}
	which says that the effective value of Newton's constant is 
	\begin{equation}
	G_{d+1} = \frac{(d-2)^{2}(d-4)^{2}\pi^{d/2-1}}{16(d-1)}\frac{\Gamma(d)}{(\Gamma(d/2))^{3}}. \label{G_scalar}
	\end{equation}
	
	\subsubsection{$d=6$} 
	
	Using \eqref{hrt_two}, we get
	\begin{equation}
	S(t_{0}) =\mbox{ }\frac{\mathcal{A}}{2G_{7}}\mbox{ }\int_{\delta}^{\infty}dz\mbox{ }\frac{h^{2}(z,t_{0})}{z^{5}}
	\end{equation} 
	After inserting \eqref{h_scalar} in the above expression, and extracting the log-divergence, we get
	\begin{equation}
	S_{\text{log}}(t_{0}) =\mbox{ }-\frac{\mathcal{A}}{G_{7}}\mbox{ }h_{2,0}(t_{0})\mbox{ }\log\delta \label{a_two_d_six}
	\end{equation}
	Using \eqref{res_a_two_d_six}, we get
	\begin{equation}
	h_{2,0}(t_{0}) =\mbox{ }-\frac{1}{20}\phi_{0,0}^{2}(t_{0}) = -\frac{1}{20}\lambda^{2}(t_{0}),
	\end{equation}
	and hence
	\begin{equation}
	S_{\text{log}}(t_{0}) = \frac{\mathcal{A}}{20G_{7}}\lambda^{2}(t_{0}) \log\delta
	\end{equation}
	This is an example of a result which depends only on Newton's constant and not any of the other parameters in $V(\Phi)$. Then it depends only on the two-point function of $\mathcal{O}$, through \eqref{def_G}. Using \eqref{G_scalar}, we get $G_{7} = 12\pi^{2}$. Then the holographic result, extrapolated to the free scalar, is
	\begin{equation}
	S_{\text{log}}(t_{0}) = \frac{\mathcal{A}}{240\pi^2 } \lambda^{2}(t_{0})\log\delta= \frac{\mathcal{A}}{960\pi^2} m^{4}(t_{0})\log\delta \label{holographic_scalar_d_six}
	\end{equation}
If $m^2(t)$ is constant, this agrees with the well-known answer for the static case \cite{misha_myers_hung}. We will see below in Section~\ref{sec-freefield} that we have full agreement also in the time-dependent case.
	
	\subsubsection{$d=8$}
	
	Using \eqref{hrt_two}, we get
	\begin{equation}
	S(t_{0}) =\mbox{ }\frac{\mathcal{A}}{2G_{9}}\mbox{ }\int_{\delta}^{\infty}dz\mbox{ }\frac{h^{3}(z,t_{0})}{z^{7}}
	\end{equation} 
	After inserting \eqref{h_scalar} in the above expression, and extracting the log-divergence, we get
	\begin{equation}
	S_{\text{log}}(t_{0}) =\mbox{ }-\frac{3\mathcal{A}}{2G_{9}}\mbox{ }\Big(h_{2,2}(t_{0}) + h_{3,0}(t_{0}) \Big)\mbox{ }\log\delta \label{a_two_d_eight}
	\end{equation}
	Using \eqref{res_a_two_d_eight} and using \eqref{dic_coupling}, we get
	\begin{equation}
	h_{2,2}(t_{0}) + h_{3,0}(t_{0})  =\mbox{ }\frac{\kappa}{126}\lambda^{3}(t_{0}) + \frac{1}{168}\dot{\lambda}^{2}(t_{0}) + \frac{1}{84}\lambda(t_{0})\ddot{\lambda}(t_{0}) ,
	\end{equation}
	and hence
	\begin{equation}
	S_{\text{log}}(t_{0}) = -\frac{\mathcal{A}}{2G_{9}}\Big( \frac{\kappa}{42}\lambda^{3}(t_{0}) + \frac{1}{56}\dot{\lambda}^{2}(t_{0}) + \frac{1}{28}\lambda(t_{0})\ddot{\lambda}(t_{0}) \Big) \log\delta
	\end{equation}
	where $G_{9} = 120\pi^{3}$ is given by \eqref{G_scalar}. Note that the coefficent of the $\lambda^{3}(t)$ term is proportional to $\kappa$ and hence it's not expected to be universal for all unperturbed CFTs. 
	
	\subsection{Fermionic mass operator: $\alpha=1$} \label{holographic_fermion}

	In this subsection, we turn to the case $\alpha=1$. This corresponds in the free field limit to a Dirac mass for a fermion:
	\begin{align}
	\mathcal{O}(x) =&\mbox{ } \bar{\psi}(x)\psi(x) \\
	\lambda(t) =&\mbox{ }m(t)
	\end{align}
The ansatz for $h(z,t)$ given in \eqref{ans_h} for $\alpha = 1$ is
	\begin{equation}
	h(z,t) =\mbox{ }1 + z^{2}h_{2,0}(t) + z^{3}h_{3,0}(t) + z^{4}\Big(h_{2,2}(t) + h_{4,0}(t)\Big) + O(z^{5}) \label{h_fermion}
	\end{equation}
By a power counting argument, we find that there is a log divergence for all $d>3$.  The absence of the log term in $d=3$ is consistent with the known result \cite{misha_myers_hung, RS1} that the change in  the entanglement entropy to first order in a perturbation away from a CFT is zero: the coefficient of the log term, if present, would have been proportional to $m$.

To determine the effective value of $G_{d+1}$ using \eqref{def_G}, we will use the following two point function of $\mathcal{O}_{\psi}(x) =\mbox{ }\bar{\psi}(x)\psi(x)$ with itself \cite{RS4}:
	\begin{equation}
	\langle \mathcal{O}_{\psi}(x)\mathcal{O}_{\psi}(0)\rangle =\mbox{ }\frac{2^{\lf d/2 \rf}}{\Omega_{d}^{2}}\frac{1}{x^{2(d-1)}} \label{two_point_function_fermion}
	\end{equation}
	
	\subsubsection{$d=4$}
	
The log divergence in \eqref{hrt_two} in $d=4$ is equal to
	\begin{equation}
	S_{\text{log}}(t_{0}) =\mbox{ }-\frac{\mathcal{A}}{2G_{5}}\mbox{ }h_{2,0}(t_{0})\mbox{ }\log\delta \label{a_one_d_four}
	\end{equation}
where $h_{2,0}(t_{0})$ is given in \eqref{res_a_one_d_four}:
	\begin{equation}
	h_{2,0}(t_{0}) =\mbox{ }-\frac{1}{12}\phi_{0,0}^{2}(t_{0})=\mbox{ }-\frac{1}{12}\lambda^{2}(t_{0}).
	\end{equation}
Using \eqref{def_G} and \eqref{two_point_function_fermion}, we get $G_{5} = \frac{\pi}{2}$.
Then we have 	
	\begin{equation}
	S_{\text{log}}(t_{0}) =\frac{\mathcal{A}}{12\pi}\lambda^2(t_{0}) \log\delta \label{fer_d4_hol}
	\end{equation}

	\subsubsection{$d=5$}
	
In this case, the log divergence in the entanglement entropy is equal to
	\begin{equation}
	S_{\text{log}}(t_{0}) = -\frac{3\mathcal{A}}{4G_{6}}\mbox{ }h_{3,0}(t_{0}) \log\delta \label{a_one_d_five}
	\end{equation}
We copy the result for $h_{3,0}(t_{0})$ from \eqref{res_a_one_d_five}:
	\begin{equation}
	h_{3,0}(t_{0}) =\mbox{ }\frac{\kappa}{36}\mbox{ }\phi_{0,0}^{3}(t_{0})
	\end{equation}
Using \eqref{def_G} and \eqref{two_point_function_fermion}, we get $G_{6} = 8$.	
Therefore, the entanglement entropy becomes
	\begin{equation}
	S_{\text{log}}(t_{0}) =-\frac{\mathcal{A}}{384} \kappa \phi^{3}_{0,0}(t_{0}) \log\delta = -\frac{\mathcal{A}}{384} \kappa \lambda^3(t_{0}) \log\delta
	\end{equation}
In this case the answer depends on the bulk interaction strength $\kappa$, and hence is not expected to be universal across all theories.

	\subsubsection{$d=6$}
	
In this case, the entanglement entropy in \eqref{hrt_two} simplifies to 
	\begin{equation}
	S(t_{0}) =\mbox{ }\frac{\mathcal{A}}{2G_{7}}\mbox{ }\int_{\delta}^{\infty}dz\mbox{ }\frac{h^{2}(z,t_{0})}{z^{5}} 
	\end{equation}
By expanding \eqref{h_fermion} in the above expression and extracting the log term, we get
	\begin{equation}
	S_{\text{log}}(t_{0}) =\mbox{ }-\frac{\mathcal{A}}{2G_{7}}\mbox{ }\Big( h_{2,0}^{2}(t_{0}) + 2h_{2,2}(t_{0}) + 2h_{4,0}(t_{0}) \Big)\mbox{ }\log\delta \label{a_one_d_six}
	\end{equation}
Using \eqref{res_a_one_d_six}, we get
	\begin{align}
	h_{2,0}^{2}(t_{0}) + 2h_{2,2}(t_{0}) + 2h_{4,0}(t_{0}) = \frac{(117-65\kappa^{2} + 45\omega)}{7200}\mbox{ }\phi_{0,0}^{4}(t_{0}) + \frac{1}{80}\mbox{ }\partial_{t}^{2}\Big(\phi_{0,0}^{2}(t_{0})\Big)
	\end{align}
Using \eqref{def_G} and \eqref{two_point_function_fermion}, we get $G_{7} = \frac{3}{2}\pi^{2}$.		
Combining these results gives us the log term divergence which not only depends on the instantaneous value of the coupling, but also on the instantaneous values of its derivatives. Notice that the derivative term does not depend on the higher-point interactions. Putting together the above results gives
\begin{equation}
S_{\partial^2, \text{log}}(t_{0}) = -\frac{\mathcal{A}}{240\pi^{2}}\partial_{t}^{2}\Big(\lambda^{2}(t_{0})\Big)\log\delta \label{fer_d6_hol}
\end{equation}

	
\section{Field Theory Calculation}\label{sec-freefield}

	In this section, we use the Hamiltonian formalism to find the time evolution of the log term in the entanglement entropy for a general field theory to first order in perturbation theory. We will identify the linear term in the time-dependent entropy with a certain correlation function in the original theory. Using the free scalar field as a guiding example, we will go on to evaluate the divergent parts of this correlation function in terms of the spectral function $c^{(0)}(\mu)$ of the theory, defined in terms of the two-point function of the trace of the energy-momentum tensor~\cite{Cappelli:1990yc}. Finally, we use this general prescription to calculate the answer for a free Dirac field.

\subsection{Entropy Perturbation as a Correlation Function}

The time dependent Hamiltonian of our theory is 
	\begin{equation}
	H(t) = H_{0} + \lambda(t)\mathcal{O},
	\end{equation}
where $H_{0}$ is the Hamiltonian of the unperturbed  theory, and $\lambda(t \le 0) = 0$.
	
We start with the system in the vacuum state, $\ket{0}$, of the unperturbed Hamiltonian. The global state at any time $t>0$ is given by the time-dependent density matrix, $\hat{\rho}(t) = \ket{\chi(t)}\bra{\chi(t)}$, where $\ket{\chi(t)}$ solves the Schr\"odinger equation generated by $H(t)$ with initial condition $\ket{0}$. The time-dependent reduced density matrix of some spatial region $B$ is given by the tracing over the Hilbert space of complementary region, $\bar B$. That is
	\begin{equation}
	\rho(t) = {\rm Tr}_{\bar B}\mbox{ }\ket{\chi(t)}\bra{\chi(t)}.
	\end{equation}
We define the time dependent modular Hamiltonian $K(t)$ of region $B$ as
	\begin{equation}
	\rho(t) \equiv\mbox{ }e^{-K(t)},
	\end{equation}
where $K(t)$ has only support in the region $B$. In other words, this operator acts as the identity operator on the Hilbert space on the region $\bar B$.
	
	The entanglement entropy of $B$ can be written using the modular Hamiltonian:
	\begin{align}
	S(t) =&\mbox{ }-{\rm Tr}_{B}\mbox{ }\rho(t)\log\rho(t)\nonumber\\
	=&\mbox{ }{\rm Tr}_{B}\Big( K(t){\rm Tr}_{\bar B}\ket{\chi(t)}\bra{\chi(t)}\Big)\nonumber\\
	=&\mbox{ }{\rm Tr}_{B\cup \bar B}\Big(K(t)\ket{\chi(t)}\bra{\chi(t)}\Big)\nonumber\\
	=&\mbox{ }\bra{\chi(t)}K(t)\ket{\chi(t)} \label{ent_mod}.
	\end{align}
To extract the term linear in $\lambda(t)$, we define the state in the interaction picture. That is
	\begin{equation}
	\ket{\tilde{\chi}(t)} \equiv e^{iH_{0} t}\ket{\chi(t)}
	\end{equation}
where the interaction picture state solves the equation
	\begin{equation}
	i\partial_{t}\ket{\tilde{\chi}(t)} = \lambda(t) \tilde{\mathcal{O}}(t) \ket{\tilde{\chi}(t)},
	\end{equation}
where $\tilde{\mathcal{O}}(t) = e^{iH_{0}t}\mathcal{O}e^{-iH_{0}t}$. We can solve the above equation perturbatively 
	\begin{equation}
	\ket{\tilde{\chi}(t)} =\ket{0} + \ket{\tilde{\chi}^{(1)}(t)} + \frac{1}{2}\ket{\tilde{\chi}^{(2)}(t)}  + \ldots
	\end{equation}
where the first order correction is
	\begin{equation}
	\ket{\tilde{\chi}^{(1)}(t)} = -i\int_{0}^{t}dt'\mbox{ }\lambda(t')\tilde{\mathcal{O}}(t')\ket{0}. \label{first_state}
	\end{equation}
We expand the modular Hamiltonian in the same fashion
	\begin{equation}
	K(t) = K_{0} + K^{(1)}(t) + \frac{1}{2} K^{(2)}(t) + \cdots,
	\end{equation}
where $K_{0}$ is the modular Hamiltonian of the vacuum state of the CFT, and hence is time independent, and $K^{(n)}$ involves $n$ powers of $\lambda$.

		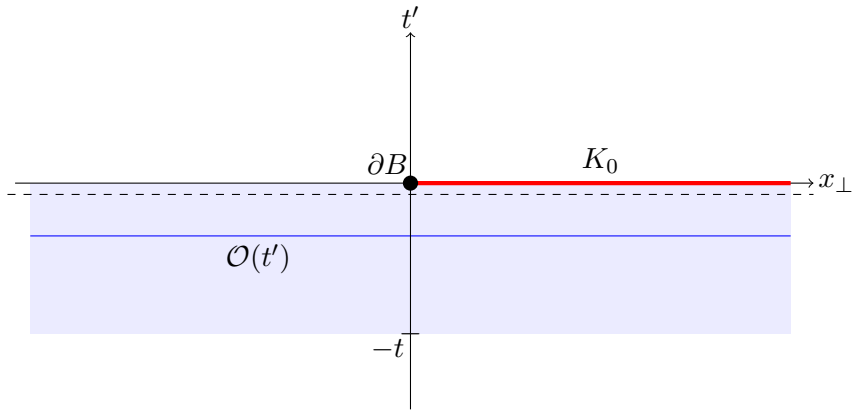
\begin{figure}
		\center
\begin{tikzpicture}
	\fill [blue!8!white] (-5,0) rectangle (5,-2);
	\draw [dashed] (-5.3,-.15) -- (5.3,-.15);
	\draw [->]  [black](-5.2,0) --(5.3,0);
	\draw [blue](-5,-0.7) --(5,-0.7);
	\draw [ultra thick, red](0,0) --(5,0);
	\draw [->]  [black](0,-3) --(0,2);
		\node at (5.6,0) {$x_{\perp}$};
		\node at (0,2.2) {$t'$};
		\node at (-.3,-2.2) {$-t$};
		\node at (0,-2) {$-$};
		\node at (2.5,0.3) {$K_{0}$};
		\node at (-2,-1) {$\mathcal{O}(t')$};
		\fill [black](0,0) circle (.1cm);
		\node at (-.3,.25) {$\partial B$};
 \end{tikzpicture}
\caption{Two dimensional cross-section of our setup illustrating equation \eqref{first_ent}. The entangling surface, $\partial B$, is at $t'=0$ and $x_{\perp}=0$ (marked as the black dot). The modular Hamiltonian of the unperturbed state lives entirely in region $B$ at $t'=0$, indicated by the red line. We first compute the correlation function of $K_{0}$ with the relevant operator, $\mathcal{O}$, inserted at the blue line, and then integrate over the shaded region. The dashed line at $t'=-\delta$ serves as the UV cutoff.}
			\label{fig-formula}
		\end{figure}

Combining these expansions we get the first order change in the entanglement entropy:
	\begin{equation}
	S^{(1)}(t) =\mbox{ }\bra{\tilde{\chi}^{(1)}(t)}e^{iH_{0}t}K_{0}\ket{0} + \bra{0}K_{0}e^{-iH_{0}t}\ket{\tilde{\chi}^{(1)}(t)} + \bra{0}K^{(1)}(t)\ket{0}. \label{first_ent_int}
	\end{equation}
The normalization of the reduced density matrix provides a constraint on the modular Hamiltonian:
	\begin{align}
	 {\rm Tr}_{B} \, e^{-K(t)} = 1.
	\end{align}
Expanding this constraint to leading order yields
	\begin{equation}
	\bra{0}K^{(1)}\ket{0} = 0,
	\end{equation}
which simplifies \eqref{first_ent_int} to
	\begin{equation}
	S^{(1)}(t) = 2{\rm Re}\left[\bra{0}K_{0}e^{-iH_{0}t}\ket{\tilde{\chi}^{(1)}(t)}\right] = -2\int_{-t}^{0}dt'\lambda(t'+t)~{\rm Re}\left[i\bra{0}K_{0}\tilde{\mathcal{O}}(t')\ket{0}\right] . 
	\label{first_ent}
	\end{equation}
This is one of our main results. The content of this formula is summarized in Fig.~\ref{fig-formula}.
	
	The formula \eqref{first_ent} is applicable for any spatial region $B$. However, the vacuum modular Hamiltonian $K_0$ is not known for a general region. We will therefore again restrict ourselves to the case where entangling surface is an infinite plane, so that the region $B$ is a half-space. In that case, the modular Hamiltonian of the vacuum state is well known \cite{Bis75,Bis76,Kabat:1994vj,Crisp07}
	\begin{equation}
	K_{0} =\mbox{ }2\pi\mbox{ }\int d\x_{\|}\int_{0}^{\infty}dx_{\perp}\mbox{ }x_{\perp}T_{00}(t=0,x_{\perp},\x_{\|}). \label{mod_hamiltonian}
	\end{equation}
Here we have adopted the following notation for spatial  vectors, $\mathbf{v}$. We write $\mathbf{v} = (v_{\perp},\mathbf{v}_{\|})$, where $v_{\perp}$ is the component of $\mathbf{v}$ in the direction orthogonal to the entangling surface and $\mathbf{v}_{\|}$ is the projection of $\mathbf{v}$ parallel to the entangling surface.

Putting this together, we see that the first-order change in the entropy is related to the commutator of the energy density and the perturbation:
	\begin{equation}
	S^{(1)}(t) = -2\pi i \int_{-t}^{0}dt'\int d\x_{\|}\int_{0}^{\infty}dx_{\perp}\mbox{ }x_{\perp} \lambda(t'+t) \bra{0} \left[T_{00}(0,x_{\perp},\x_{\|}),\tilde{\mathcal{O}}(t')\right]\ket{0}. \label{eq-entropycomm}
	\end{equation}
Notice that the interaction picture operator $\tilde{\mathcal{O}}(t)$ is identical to the Heisenberg operator $\mathcal{O}(t)$ in the original $\lambda(t) =0$ theory, i.e., when the total Hamiltonian is equal to $H_0$. In the following, $\mathcal{O}$ itself will be the integral of a local operator, so we effectively reduce the problem to a commutator of local operators (or equivalently the imaginary part of a two-point function) in the original theory.

	
\subsection{Example: free scalar field} 
\label{free_scalar_section}

In the static perturbation theory the first order change in the entanglement entropy after perturbing a CFT by a relevant operator vanishes \cite{RS1}. It follows from \reef{eq-entropycomm} that similar conclusion holds in the time dependent case. To resolve this difficulty, we will start with the free scalar with non-zero mass, $m$, and then introduce a time dependent perturbation to the mass, $m^2(t) = m^2 + 2\lambda(t)$.\footnote{The factor of two is so our definition of $\lambda$ remains consistent with the previous section.} The action of our theory is\footnote{We use mostly minus signature.}
	\begin{equation}
	I[\phi(t,\x)] =\mbox{ }\int dtd\x\mbox{ } \Big(\frac{1}{2}(\partial\phi(t,\x))^{2} - \frac{1}{2}m^{2}\phi^{2}(t,\x) - \lambda(t)\phi^{2}(t,\x)\Big)
	\end{equation}
where $\lambda(t \le 0) = 0$. The perturbing operator is $\mathcal{O}(t') = \int d\x' \, \phi^{2}(t',\x')$, where the time dependence of these operators is governed by the free Hamiltonian with $\lambda=0$. From \eqref{eq-entropycomm} we see that the entropy is determined by correlation functions of $T_{00}$ and $\phi^2$. The energy-momentum tensor of the free scalar field is given by
	\begin{equation}
	T_{\mu\nu} = \partial_{\mu}\phi\partial_{\nu}\phi - \frac{1}{2}\eta_{\mu\nu}\Big( (\partial\phi)^{2} - m^{2}\phi^{2}\Big) + \xi \Big( \eta_{\mu\nu}\partial^{2} - \partial_{\mu}\partial_{\nu}\Big)\phi^{2} \label{scalar_tab}
	\end{equation}
where the last term is a possible improvement term. The energy momentum tensor is traceless if it is massless and $\xi = \xi_{c} \equiv\mbox{ }\frac{d-2}{4(d-1)}$, in which case the scalar field is said to be conformally coupled. The other noteworthy case is the minimally coupled scalar, where $\xi=0$. The required correlation functions are easily computed using Wick's theorem:
\be \label{eq-Tphi2}
\bra{0}  T_{00}(x)  \phi^2(x')\ket{0} = (\partial_t G)^2 + (\nabla G)^2 + m^2 G^2 - 2\xi \nabla^2G^2,
\ee
where
\be \label{eq-phiphi}
G(x-x') = \bra{0}  \phi(x) \phi(x')\ket{0} = \int \frac{d\p}{(2\pi)^{d-1}} \frac{1}{2E_\p}e^{-ip\cdot(x-x')}
\ee
is the (unordered Lorentzian) two-point function of the scalar field. The expectation value of the commutator $\left[  T_{00}(x), \phi(x')^2\right]$ is simply twice the imaginary part of the correlation function, up to a factor of $i$. Before taking the imaginary part, we first perform the required integrations, the details of which we leave to Appendix~\ref{app-integrations}. From \eqref{eq-answer} we find
\begin{align}
\int_{-t}^{0}dt'\int d\x' \int d\x_{\|}&\int_{0}^{\infty}dx_{\perp}\mbox{ }x_{\perp} \lambda(t'+t) \bra{0}  T_{00}(0,x_{\perp},\x_{\|}) {\phi}^{2}(t',\x')\ket{0}\nonumber\\
  &= \mathcal{A}\int_{0}^{t}dt'\, \lambda(t')\int \frac{d\p}{(2\pi)^{d-1}} \left(\frac{1 - p_{\perp}^2  /E^2 - 4\xi }{8E^2}\right)e^{2iE(t'-t)}. 
\end{align}
Using this result in \eqref{eq-entropycomm} yields
	\begin{align}
	S^{(1)}_\mt{scalar}(t) =& \,\frac{\Omega_{d-1}}{4(2\pi)^{d-2}} \, \mathcal{A}\int_{0}^{t}dt' \lambda(t') \int_{0}^{\infty}dp\mbox{ }\Big( 4(\xi_{c}-\xi)\mbox{ }\frac{p^{d-2}}{E^{2}}\mbox{ } +\mbox{ } \frac{m^{2}}{d-1}\mbox{ }\frac{p^{d-2}}{E^{4}}\Big)\sin\big(2E(t' - t)\big) 
	\end{align}
where $\Omega_{d} = \frac{2\pi^{d/2}}{\Gamma(d/2)}$.

To isolate UV divergences in the above expression, we expand the integrand around $x\equiv E/m\gg 1$ 
\bea
 S^{(1)}_\mt{scalar}(t) &=& \frac{\Omega_{d-1}}{4(2\pi)^{d-2}(d-1)} m^{d-3}\mathcal{A}  
 \sum_{n=0}^\infty  {(-1)^n \Gamma\({d-1\over 2}\) \over \Gamma\({d-1 - 2n\over 2} \)\Gamma(n+1)} \int_{0}^{t}dt' \lambda(t')
  \non
 &\times& \int_{1}^{\infty} dx \, x^{d-6-2n} \Big(4(\xi_{c}-\xi)(d-1)x^2 + 1\Big)\sin\big( 2(t'-t) m \, x  \big) ~.
\eea
Next we note that for any given $d$, the integral over $x$ in the above expression results in a UV-divergent $S^{(1)}$ if and only if $d-4-2n\geq 0$. There are only a finite number of such terms, and therefore we regularize them by assuming that $n$ is sufficiently large, carry out the integral over $x$ and treat the special values of $n$ by analytic continuation. 

The final result depends on the parity of $d$. For odd $d$ there are no divergences, and $S^{(1)}$ is finite.\footnote{The non universal divergences vanish for our choice of regularization scheme.} On the other hand, for even $d$ we obtain
\bea
 S^{(1)}_\mt{scalar}(t) &=& \frac{\Omega_{d-1}}{4(2\pi)^{d-2}(d-1)} m^{d-3}\mathcal{A}  
 \sum_{0\leq n\leq {d-4\over2}}  {(-1)^{d\over 2} \Gamma\({d-1\over 2}\) \over \Gamma\({d-1 - 2n\over 2} \)\Gamma(n+1)} \int_{0}^{t}dt' \lambda(t')
  \\
 &\times&\( {4(\xi_{c}-\xi)(d-1)  \Gamma(d-2n-3)\over \big(2(t'-t)m\big)^{d-2n-3} } 
 - { \Gamma(d-2n-5)\over \big(2(t'-t)m\big)^{d-2n-5}}\) \, + \, \text{finite terms} ~.
 \nonumber
\eea
Introducing a UV cut off surface $t'=t-\delta$ to ensures that the integral over $t'$ is finite, and expanding $\lambda(t')$ in the vicinity of the cutoff yields
\bea
 S^{(1)}_\mt{scalar}(t) &=& \frac{(-1)^{d\over 2}}{(16\pi)^{d-3\over 2}\Gamma\({d-1\over 2}\)} \mathcal{A}  \(  (\xi_c-\xi) \lambda^{(d-4)}(t)  -   
 \sum_{0\leq n\leq {d-6\over2}}  {(4m^2)^{n} \Gamma\({d-1\over 2}\)\over \Gamma\({d-1\over 2} - n \)\Gamma(n+1)}  \right.
  \non
  &\times& \left. m^2\,\lambda^{(d-2n-6)}(t)\Big( {2(d-2n-3)\over n+1}(\xi - \xi_{c}) + {1\over d-1}\Big) \) \log\delta + \, \text{finite terms} ~.
 \non
 \label{scalar_entropy_final}
\eea

There are two points to make about this result: (i) It gives us the expected adiabatic expansion of the time-dependence of the divergent terms in the entanglement entropy, where in accord with local nature of the divergences only the instantaneous value of the coupling and its derivatives matter.  (ii) Only even numbers of derivatives appear in the divergent terms, which is consistent with the holographic result.  

As an initial check of our calculations, let us consider a static mass perturbation, i.e. $\lambda(t) = \lambda$. In this case only $n=(d-6)/2$ term in \eqref{scalar_entropy_final} survives, and we get
\begin{equation}
  S^{(1)}_\mt{scalar} = (1-6\xi)\,\frac{(-1)^{\frac{d}{2}}\mathcal{A}}{3(4\pi)^{\frac{d-2}{2}} \Gamma\({d-2\over 2}\)}\mbox{ }m^{d-4}\lambda\mbox{ }\log\delta + \, \text{finite terms} ~.
\end{equation}
For $\xi=0$ this formula matches the well-known first-order expansion of the static result \cite{Hertzberg:2010uv,Huerta:2011qi,Lewkowycz:2012qr}, while for $\xi\neq 0$ it agrees with \cite{Casini:2014yca,Ben-Ami:2015zsa,Akers:2015bgh} .

For the time-dependent $\lambda(t)$ in $d=4,6,8$, the explicit form of the first order correction is 
\bea
S^{(1)}_\mt{scalar}(t)\Big|_{d=4} &=& (\xi_c - \xi)\frac{\mathcal{A}}{2\pi}\mbox{ }\lambda(t)\log\delta + \, \text{finite terms} ~.
\\
 S^{(1)}_\mt{scalar}(t)\Big|_{d=6} &=& \frac{-\mathcal{A}}{48\pi^{2}}\Big( (\xi_c - \xi)\ddot{\lambda}(t)+\left(6(\xi_c-\xi)-{1\over 5}\right)m^{2}\lambda(t)\Big)\log\delta
 + \, \text{finite terms} ~.
 \non
 S^{(1)}_\mt{scalar}(t)\Big|_{d=8} &=& \frac{\mathcal{A}}{1920\pi^{3}}\Big( (\xi_c - \xi)\lambda^{(4)}(t)
 +\left(10(\xi_c-\xi)-\frac{1}{7}\right)m^2 \ddot{\lambda}(t)
 \non
 &&\quad\quad\quad\quad + 10\left(3(\xi_c-\xi)-\frac{1}{7}\right)m^4 \lambda(t) \Big)\log\delta + \, \text{finite terms} ~.
 \nonumber
\eea
Setting $\xi=\xi_c$ we find that terms proportional to $m^2\lambda$ and $m^2\ddot\lambda$ in $d=6$ and $d=8$ respectively agree with their holographic counterparts.\footnote{To see matching recall that $m^2(t) = m^2 + 2\lambda(t)$.} Furthermore, linear terms in $\lambda(t)$ and its derivatives vanish in all dimensions which is also in full agreement with holography. Such a match between the calculations for the free field theory and strongly coupled $\mathcal{N}=4$ SYM might seem surprising at first sight. However, as we argue in the next section, similarly to the time independent case \cite{misha_myers_hung,Liu:2012eea}, this match can be attributed to the universality of $\langle T_{\mu\nu}O\rangle$, $\langle OO\rangle$ and $\langle T_{\mu\nu}OO\rangle$ in a conformal field theory \cite{RS4,Faulkner:2014jva}. We note that tuning $\xi$ to a conformal value $\xi_c$ was important to ensure tracelessness of the stress tensor in a CFT.

\subsection{Spectral representation}
\label{sec:specrep}

The scalar field theory example of the previous section suggests that the time dependent universal terms of EE can be easily identified if we rewrite \reef{eq-entropycomm} as an integral over energy states. The main goal of this section is to perform such spectral analysis for an interacting QFT which is driven by the action with only one operator, $\mathcal{O}(x)$, of scaling dimension $\Delta<d$. The corresponding  relevant coupling will be denoted by $\lambda$. We assume that $\lambda$ is equal to some fixed value plus a time-dependent perturbation: $\lambda(t) = \lambda_{0} + \delta\lambda(t)$.

We start from noting that our main result \reef{first_ent} (or equivalently, \reef{eq-entropycomm} in the case of planar entangling surface) can be written as follows 
\begin{equation}
 S^{(1)}(t) = -2\int_{-t}^0 dt' \lambda(t'+t) \, {\rm Re}\left[i \langle 0| K_{\lambda_{0}} \mathcal{O}(t') |0\rangle_{\lambda_{0}} \right] ~.
 \label{main2}
\end{equation}
where $\langle ... \rangle_{\lambda_{0}}$ denotes the correlator in the unperturbed theory with coupling $\lambda_{0}$.

Recalling now that by definition $K_{\lambda_{0}}$ is sitting at the $t=0$ slice, we deduce that the above correlator can be easily obtained from its Euclidean counterpart by analytic continuation. One simply evaluates the Euclidean correlator $\langle K_{\lambda_{0}} \mathcal{O}(t'_\mt{E}) \rangle_{\lambda_{0}} $ and then substitutes Euclidean time, $t'_\mt{E}$, with $i(t'+i\epsilon)$. 

Moreover, if we implement conformal perturbation theory and expand this correlator around $\lambda_{0} = 0$, we will get \cite{RS4} 
\begin{equation}
\langle 0| K_{\lambda_{0}} \mathcal{O} |0\rangle_{\lambda_{0}} = -\lambda_{0}\Big( \langle K_{0}\mathcal{O}\mathcal{O} \rangle_{0} - \langle \mathcal{O}\mathcal{O} \rangle_{0} \Big) + O(\lambda_{0}^{2}) \, . \label{con-pert-linear}
\end{equation}
where we have used the fact that the correlator $\langle K_{0}\mathcal{O} \rangle_{0}$ vanishes for the CFT \footnote{This just means that the first order change in the entanglement entropy after perturbing a CFT by a relevant operator vanishes. \cite{RS1}}. This observation implies that the result for $S^{(1)}(t)$ to linear order in $\lambda_{0}$ will be independent of the underlying CFT.\footnote{In fact, this conclusion also holds in the presence of multiple relevant couplings in the theory.} This follows from the universality of conformal correlators $\langle OO\rangle$ and $\langle T_{\mu\nu}OO\rangle$ \cite{Osborn:1993cr,Erdmenger:1996yc}. Thus we conclude that all terms of $S^{(1)}(t)$ which are proportional to $\lambda_{0} \, \delta\lambda(t)$ (or $\lambda_{0} \, \partial^{2}_{t}\delta\lambda(t)$ and so on) are identical for {\it any} perturbed CFT. In particular, this explains the perfect matching, mentioned in the previous section, between certain terms in the holographic and free field theory calculations. This can be seen in the following way: Note that the terms proportional to $\lambda_{0} \, \delta\lambda(t)$ in $S^{(1)}(t)$ complete into terms which are quadratic in $\lambda(t)$ or its derivatives in the full time dependent entropy, $S(t)$. Recall from the section-\ref{sec-holo} that the terms that are quadratic in $\lambda(t)$ or its derivatives ({\it eg.} $\lambda^{2}(t)$ or $\lambda(t)\ddot{\lambda}(t)$) in the holographic entanglement entropy only multiply the effective Newton constant, $G_{d+1}$, whereas $\lambda^{3}(t)$ contributions also multiply $\kappa$ and $\lambda^{4}(t)$ also multiply $\omega$, where $\kappa$ and $\omega$ are coupling constants in \eqref{bulk_action}. Since the effective Newton constant is determined by the boundary two point function according to \eqref{def_G}, the coefficients of terms quadratic in $\lambda(t)$ or its derivatives in the holographic calculation are therefore universal. See~\cite{RS4,Faulkner:2014jva} where this universality was observed in a static setup.  

By assumption there is only one relevant coupling, $g$, in the action. Hence, it can be shown that \cite{Smolkin:2014hba,RS3,Casini:2014yca}
\be 
 \langle \,  \mathcal{O}(t'_\mt{E},x_{\perp},\x_{\|}) \,  K_0 \,\rangle =
 {  \Omega_d \over g\,(d-\Delta + \beta) 2^{d-1} (d-1)(d+1) \Gamma(d) }  \int_0^\infty d \mu ~ \mu^2 \, c^{(0)}(\mu) \, K_0\big(\mu\, \sqrt{t^2_\mt{E}+x_\perp^2}\big)~,
\ee
where $c^{(0)}(\mu)$ is the spectral function defined by the 2-point function of the energy-momentum trace \cite{Cappelli:1990yc}, $K_0$ on the right-hand side is the modified Bessel function of the second kind\footnote{On the left-hand side, as in all other equations, $K_0$ is the unperturbed modular Hamiltonian.}, $\beta$ is  the anomalous dimension of $g$ and $\Omega_d=2\pi^{d/2}/\Gamma(d/2)$.

Like in the free scalar case we first integrate this correlator over $x_\|$ and $x_\perp$, 
\be 
 \int d^{d-2}x_\| \int dx_\perp \langle \,  \mathcal{O}(x) \,  K_0 \,\rangle =
 {  \pi \, \Omega_d \, \mathcal{A} \over g\,(d-\Delta + \beta) 2^{d-1} (d-1)(d+1) \Gamma(d) }  \int_0^\infty d \mu ~ \mu \, c^{(0)}(\mu) \, e^{-|t_\mt{E}|\mu}~.
\ee
Next, we analytically continue this result back to Lorentzian signature $t_\mt{E}\to i(t'+i\epsilon)$, substitute it into \reef{main2} and get the following spectral representation of the general formula \reef{eq-entropycomm}
\begin{equation}
 S^{(1)}(t) =  {  2 \pi \, \Omega_d \, \mathcal{A} \over g\,(d-\Delta + \beta) 2^{d-1} (d-1)(d+1) \Gamma(d) }
  \int_0^\infty d \mu ~ \mu \, c^{(0)}(\mu) \, \Lambda(t,\mu)~. 
 \label{specrep}
\end{equation}
where we have absorbed the time dependence in the function $\Lambda(t,\mu)$ defined as
\begin{equation}
 \Lambda(t,\mu) =\mbox{ } \int_{-t}^0 dt' \lambda(t'+t) \, \sin(t'\mu)~. 
 \label{specrep_time}
\end{equation}
The divergences of \eqref{specrep} depend on the $\mu\to \infty$ behavior of $c^{(0)}(\mu)$. Expanding $c^{(0)}(\mu)$ near $\mu = \infty$ results in a finite number of terms of the form $\mu^s$ for $s >0$, and upon substituting these terms into \eqref{specrep} we find integrals of the form
\be\label{eq-divergences}
 \int_0^\infty d \mu \int_{-t}^0 dt'~ \mu^{s+1} \lambda(t'+t) \, \sin(t'\mu).
\ee
Note that the oscillatory nature of the integrand means that integrating over $\mu$ alone is not enough to give a divergence: integrating over $\mu$ would only produce a factor $\propto |t'|^{-s-2}$, but the subsequent integral over $t'$ could have a divergence near $t'=0$, depending on the behavior of $\lambda(t+t')$.

To isolate the divergent terms, we begin by integrating \eqref{specrep_time} by parts with respect to $t'$ a total of $2N+1$ times. This results in the identity
\begin{align}
 \int_{-t}^0 dt'~ \lambda(t'+t) \, \sin(t'\mu) &=  -\frac{\lambda(t)}{\mu} + \frac{\lambda^{(2)}(t)}{\mu^{3}} +\cdots -(-1)^N \frac{\lambda^{(2N)}(t)}{\mu^{2N+1}}   \nonumber\\
 &+ \sin(\mu t)\left( \frac{\lambda^{(1)}(0)}{\mu^{2}} - \frac{\lambda^{(3)}(0)}{\mu^{4}} +\cdots -(-1)^N \frac{\lambda^{(2N-1)}(0)}{\mu^{2N}}  \right) \nonumber\\
 &+ \cos(\mu t) \left(\frac{\lambda(0)}{\mu}-  \frac{\lambda^{(2)}(0)}{\mu^{3}} +\cdots +(-1)^N \frac{\lambda^{(2N)}(0)}{\mu^{2N+1}} \right)\nonumber\\
 &- (-1)^N\int_{-t}^0 dt'~ \mu^{-2N-1} \lambda^{(2N+1)}(t'+t) \, \sin(t'\mu).
 \end{align}
If $2N>s+1$, then the last line produces no UV divergences upon substitution back into \eqref{eq-divergences}. The oscillatory terms in the middle two lines likewise produce no divergences in \eqref{eq-divergences}. All that remains is the first line. There we see clearly that a logarithmic divergences is produced if and only if $s$ is an odd integer, with the coefficient of the divergence proportional to $\lambda^{(s+1)}(t)$. The non-universal coefficients can also be determined this way.

Thus we learn that the quasi-adiabatic expansion holds for the divergences in the entanglement entropy in a general interacting theory to this order in perturbations, and that the coefficients are entirely determined by the large-$\mu$ behavior of the spectral function $c^{(0)}(\mu)$.

\subsection{Example: free Dirac field}

Let us implement \reef{specrep} in the case of free massive Dirac field with a time-dependent mass deformation. In this case $\Delta=d-1$, $\bt=0$, and the spectral function is given by \cite{Cappelli:1990yc} 
\be
 c^{(0)}_\mt{Dirac}(\mu)=2^{[d/2]}\, {2(d+1)(d-1)\over \Omega_d^2} \, m^2 \, \mu^{d-5}\(1-{4m^2\over \mu^2}\)^{(d-1)/2}\Theta(\mu-2m)~.
 \label{cDirac}
\ee
Or equivalently, 
\be
  c^{(0)}_\mt{Dirac}(\mu)=2^{[d/2]}\, {2(d+1)(d-1)\over \Omega_d^2} \, m^2  \, \mu^{d-5}\Theta(\mu-2m)  \sum_{n=0}^\infty 
  {(-1)^n \Gamma\({d+1\over 2}\) \over \Gamma\({d+1 - 2n\over 2} \)\Gamma(n+1)}  \({2m\over \mu}\)^{2n} ~.
\ee
This representation is useful for analyzing the UV divergences, which are built from the modes satisfying $\mu/(2m)\gg 1$. 

Thus, for the free Dirac fermion we get
\bea
 S^{(1)}_\mt{Dirac}(t) &=&  {    2^{[d/2]} \pi \over \,\Omega_d \Gamma(d) } \, m^{d-2} \mathcal{A} 
 \sum_{n=0}^\infty {(-1)^n \Gamma\({d+1\over 2}\) \over \Gamma\({d+1 - 2n\over 2} \)\Gamma(n+1)}
 \int_{-t}^0 dt' \lambda(t'+t) \, 
 \non
 &\times&\int_1^{\infty} dx \, x^{d-4-2n} \sin\big(2mt' x\big) ~,
\eea
where we introduced a dimensionless integration variable $x=\mu/(2m)$. From the above expression, it is obvious that only terms satisfying $d-4-2n\geq 0$ result in UV divergences, whereas for $d-4-2n < 0$ integrals over $t'$ and $x$ are finite. 

In fact, the universal divergences can be readily isolated. We assume that $n$ is sufficiently large and carry out integral over $x$. Small values of $n$ are treated by analytic continuation. Hence,
\bea
 S^{(1)}_\mt{Dirac}(t) &=&  {    2^{[d/2]} \pi \over \,\Omega_d \Gamma(d) } \, m^{d-2} \mathcal{A} 
 \sum_{0\leq n\leq {d-4\over 2}} {(-1)^n \Gamma\({d+1\over 2}\) \Gamma(d-3-2n)\cos\big((d-2n){\pi\over2}\big)\over \Gamma\({d+1 - 2n\over 2} \)\Gamma(n+1)}
 \non
 &\times& \int_{-t}^{-\delta} dt' {\lambda(t'+t) \over (2m t')^{d-2n-3}} ~ + ~ \text{finite terms}~.
\eea
In particular, there are no universal divergences for odd $d$,\footnote{Recall that non-universal divergences vanish within our choice of regularization scheme.} while for even $d$ the above formula reads 
\be
 S^{(1)}_\mt{Dirac}(t) =  {   2^{6-d\over 2} \pi \over \,\Omega_d \Gamma(d) } \, \mathcal{A} 
 \sum_{0\leq n\leq {d-4\over 2}} {4^{n} (-1)^{d\over 2} \Gamma\({d+1\over 2}\) m^{2n+1} \over  \Gamma\({d+1\over 2} -n \)\Gamma(n+1)}\, \lambda^{(d-2n-4)}(t) \log\delta
 + ~ \text{finite terms}~.
\ee
For a static perturbation $\lambda(t) = \lambda$, only $n=(d-4)/2$ contributes
\begin{align}
S^{(1)}_\mt{Dirac}(t)= \frac{(-1)^{\frac{d}{2}}}{3(2\pi)^{\frac{d-2}{2}}\Gamma\({d-2\over 2}\)} \, m^{d-3} \lambda\, \mathcal{A} \log\delta 
+ ~ \text{finite terms}~.
\end{align}
This result is in full agreement with known calculations in the literature \cite{Hertzberg:2010uv,Huerta:2011qi,Lewkowycz:2012qr}. For the time-dependent case, we find
\bea
S^{(1)}_\mt{Dirac}(t)\Big|_{d=4} &=& \frac{\mathcal{A}\, m \, \lambda(t)}{6\pi} \, \log\delta + ~ \text{finite terms}~,
\label{fer_d4_ft}
\\
S_\mt{Dirac}^{(1)}(t)\Big|_{d=6} &=& -\frac{\mathcal{A}\,m}{120\pi^{2}}\Big( \ddot{\lambda}(t) + 10\,m^2\lambda(t)  \Big)\log\delta 
+ ~ \text{finite terms}~.
\label{fer_d6_ft}
\eea
Note that if we take the holographic result in $d=4$, \eqref{fer_d4_hol}, and set $m(t) = m + \lambda(t)$, then we find agreement with \eqref{fer_d4_ft}. Similarly, the double derivative term in $d=6$ in \eqref{fer_d6_hol} matches with that in \eqref{fer_d6_ft}. The precise match between the calculations for the free field theory and strongly coupled holographic CFT has to do with universality of certain correlators that we discussed in the previous section, see also \cite{RS4,Faulkner:2014jva}.


\section{Conclusions}
\label{sec-discuss}

In this work we studied the evolution of the `area law' of the spatial entanglement when the relevant coupling of the field theory undergoes a quantum quench. Our main results, \reef{first_ent} and \reef{eq-entropycomm}, for the time dependent first order correction to the area law were derived under the assumption that the relevant coupling satisfies $\delta\lambda(t)/\lambda_0\ll 1$, where $\lambda_0$ is the unperturbed value and $\delta\lambda(t)$ is a time-dependent perturbation which vanishes in the past and asymptotes to a constant value in the future. 

We argued that if $\lambda_0$ is the only relevant parameter in the system, then further simplification takes place and the final expression \reef{specrep} for the time dependent correction to the area law can be expressed in terms of the spectral function, $c^{(0)}(\mu)$, which is defined in terms of the two-point correlation function of the trace of the energy-momentum tensor in the theory. 

As an application of these general formulas, we scrutinized the time dependence of the area law in a generic QFT. In particular, we illustrated that `area law' divergences reveal a quasi-adiabatic behavior, \ie they are local and depend on the instantaneous value of the relevant coupling and finitely many derivatives.  We showed that this qualitative behavior is in full agreement with the holographic prediction based on the HRT proposal \cite{HRT}. 

We also carried out explicit calculations of the universal terms in the case of free fields perturbed by a time-dependent mass, and found {\it quantitative} agreement with the HRT predictions of the coefficient of $\lambda^{2}(t)$ in the entanglement entropy.\footnote{For similar match in the static case see \cite{RS4,Faulkner:2014jva}.} From the point of view of the holographic calculation, we observed in section-\ref{sec-holo} that the coefficients of the terms which are quadratic in $\lambda(t)$ or its derivatives ( {\it eg.} $\lambda^{2}(t)$ or $\lambda(t)\ddot{\lambda}(t)$) in the entanglement entropy only multiply the effective Newton's constant, which is fixed by the boundary two point functions. Since CFT two point functions have a universal form, this ensures that coefficients of these terms are independent of the exact nature of the unperturbed CFT. From the purely field-theoretic point of view, we observed in equation \eqref{con-pert-linear} that the coefficient of $\lambda_{0} \, \delta\lambda(t)$ in $S^{(1)}(t)$ is fixed by the universal CFT correlation functions $\langle T_{\mu\nu} \mathcal{O}\mathcal{O}\rangle$ and $\langle \mathcal{O}\mathcal{O}\rangle$. As the term proportional to $\lambda_{0}\delta\lambda(t)$ in $S^{(1)}(t)$ completes into the terms quadratic in $\lambda(t)$ and its derivative, this explains the matching of the coefficients of these quadratic terms between free field theory and holographic calculations.

Up to a set of mild assumptions, the RT proposal \cite{RT} for holographic entanglement entropy in the static case was derived in \cite{Lewkowycz:2013nqa}; see also \cite{Fursaev:2006ih,Headrick:2010zt}. The time-dependent case is more complicated: we view the match between the time-dependent terms in holography and QFT as non-trivial evidence in favor of the time-dependent HRT proposal for holographic entanglement entropy.\footnote{After the completion of this work, \cite{Dong:2016hjy} appeared which presents and argument for the HRT proposal in a similar spirit to the earlier arguments for the RT proposal.}


It would be interesting to extend our analysis to the next-to-leading order correction in $\delta\lambda(t)/\lambda_0\ll 1$. Such an extension might shed light on the fundamentals of the linear growth of entanglement entropy sufficiently far from the instant of quench \cite{CC-particle,Balasubramania_one, Balasubramanian_two, Maldacena-Hartman,HL-one, HL-two,   tsunami, Asplund, CLM, Rangamani}, and help to uncover the underlying reason for the universality of this behaviour. To make a step in this direction, it is necessary to establish a consistent perturbative expansion for the modular Hamiltonian. In fact, this problem is interesting by itself since it has many valuable practical applications beyond analysis of the time-dependent program studied in this work, and we plan to address it in the future.


\section*{Acknowledgements}

We would like to thank J.~Cardy for helpful discussions. 
This work is supported in part by the Berkeley Center for Theoretical Physics, by the National Science Foundation (award numbers 1214644, 1316783, and 1521446), by fqxi grant RFP3-1323, and by the US Department of Energy under Contract DE-AC02-05CH11231. The work of M.~M. is supported in part by a Berkeley Connect fellowship.


\appendix	
	
	\section{Details of the holographic calculation}
	
	\subsection{HRT surface calculation} \label{app-HRT}
	
	In this appendix, we will solve for the position of the HRT surface and verify the claim that we made in \eqref{asym_geodesic} that at the lowest order, the solution of the stationary surface is
	\begin{equation}
	t(z) =\mbox{ }t_{0} + z^{2+2\alpha}\mbox{ }t_{2+2\alpha} + \ldots, 
	\end{equation}
	
	As mentioned in the Section~\ref{HRT_surface}, the HRT surface is given by $t=t(z)$ and $x_{1}=0$. The equation of motion for $t=t(z)$ is obtained by varying
	\begin{equation}
	A \propto \int dz \frac{h^{\frac{d-2}{2}}(z,t(z))}{z^{d-1}}\mbox{ }\sqrt{1 - f(z,t(z)) (t'(z))^{2}}.
	\end{equation}
	We find
	\begin{align}\label{eq-tmotion}
	0 =&\mbox{ } -z\Big((d-2)\dot{h} + ht'(t'\dot{f} + 2f')\Big) + fh\Big(2(d-1)t' -2zt'' +zt'^{3}f'\Big) + (d-2)zfht'(t'\dot{h}-h')\nonumber\\ +&\mbox{ } f^{2}t'^{3}\Big(-2(d-1)h + (d-2)zh'\Big)
	\end{align}
	where dot represents derivative with respect to $t$ and prime represents derivative with respect to $z$.
	
	The ansatz \eqref{ans_h}-\eqref{ans_f} for the metric functions, $h(z,t)$ and $f(z,t)$, at leading order gives
	\begin{align}
	h(z,t) =& 1 + z^{2\alpha}h_{2,0}(t)\\
	f(z,t) =& 1 + z^{2\alpha}f_{2,0}(t)
	\end{align}
	
	Lets make the ansatz $t(z) = t_{0} + z^{\beta}t_{\beta}$. The first term of the equation of motion \eqref{eq-tmotion} is proportional to $z^{2\alpha + 1}$. Out of all the other terms, the term with the smallest power of $z$ is proportional to $z^{\beta - 1}$. Thus we learn $\beta = 2+2\alpha$.
	
	However, the coefficient of the $z^{\beta-1}$ term is equal to $2\beta (d-\beta)t_{\beta}$. Therefore, the above result is only true if $d \ne \beta$. For $d = \beta$, we would have to modify the ansatz to be $t(z) = t_{0} + z^{\beta}\log(z) t_{\beta}$. We do not consider this case in the main argument of the paper, so it can be ignored.
	
	\subsection{Perturbative Solution of Einstein-scalar equations}
	
	In this section, we will find the time-dependent coefficients $h_{m,n}(t)$, $f_{m,n}(t)$ and $\phi_{m,n}(t)$ in terms of $\phi_{0,0}(t) = \lambda(t)$ by substituting the ansatz \eqref{ans_h}-\eqref{ans_p} in the Einstein-scalar equations \eqref{equation_two}-\eqref{equation_four}. We use this result in the holographic calculation of the entanglement entropy in Sections \ref{holographic_scalar}-\ref{holographic_fermion}.

	\begin{itemize}
	\item  {\bf $\alpha = 2$ and $d=6$}
	
	This case corresponds to the holographic scalar in $d=6$. From \eqref{a_two_d_six}, we know that the universal part of the time dependent entropy only depends on the coefficient $h_{2,0}(t)$. This can easily be found using \eqref{equation_two}. Expanding this equation gives us:
	\begin{equation}
	0 =\mbox{ }(10 \dot{h}_{2,0} + \phi_{0,0}\dot{\phi}_{0,0})z^{3} + O(z^{5})
	\end{equation}
	Integrating this equation, we find 
	\begin{equation}
	h_{2,0}(t) =\mbox{ }-\frac{1}{20}\phi_{0,0}^{2}(t). \label{res_a_two_d_six}
	\end{equation}
	
	\item  {\bf $\alpha = 2$ and $d=8$}
	
	To find the entanglement entropy for the holographic scalar in $d=8$ using \eqref{a_two_d_eight}, we need to find the sum: $h_{2,2}(t) + h_{3,0}(t)$. To find this, we will substitute the ansatz \eqref{ans_h}-\eqref{ans_p} into \eqref{equation_two} and \eqref{equation_four}. Expanding \eqref{equation_two} gives us
	\begin{equation}
	0 =\mbox{ }(14 \dot{h}_{2,0} + \phi_{0,0}\dot{\phi}_{0,0})z^{3} + \big(21(\dot{h}_{2,2}+\dot{h}_{3,0}) +2\dot{\phi}_{0,0}(\phi_{0,2}+\phi_{2,0}) + \phi_{0,0}(\dot{\phi}_{0,2}+\dot{\phi}_{2,0}) \big)z^{5} +  O(z^{7}), \label{a_2_d_8_ex_1}
	\end{equation}
	and expanding \eqref{equation_four} yields
	\begin{equation}
	0 =\mbox{ }\big(4(\phi_{0,2} +\phi_{2,0}) + \frac{\kappa}{2}\phi_{0,0}^{2} + \ddot{\phi}_{0,0} \big)z^{4} + O(z^{6}). \label{a_2_d_8_ex_2}
	\end{equation}
	By solving the above equations, we find
		\begin{equation}
	h_{2,2}(t) + h_{3,0}(t)  =\mbox{ }\frac{\kappa}{126}\phi_{0,0}^{3}(t) + \frac{1}{168}\dot{\phi}_{0,0}^{2}(t) + \frac{1}{84}\phi_{0,0}(t)\ddot{\phi}_{0,0}(t). \label{res_a_two_d_eight}
	\end{equation}
	
	\item {\bf $\alpha = 1$ and $d=4$}
	
	This case corresponds to the fermionic mass operator in $d=4$. The entanglement entropy in this case is given by \eqref{a_one_d_four}. To find the entropy, we need to know $h_{2,0}(t)$. Expanding \eqref{equation_two} yields
	\begin{equation}
	0 =\mbox{ }(\dot{h}_{2,0} + \frac{1}{6}\phi_{0,0}\dot{\phi}_{0,0})z + O(z^{2}).
	\end{equation}
	Integrating this equation gives
	\begin{equation}
	h_{2,0}(t) =\mbox{ }-\frac{1}{12}\phi_{0,0}^{2}(t). \label{res_a_one_d_four}
	\end{equation}
	
	\item {\bf $\alpha = 1$ and $d=5$}
	
	In this case, the entanglement entropy is given by \eqref{a_one_d_five}, and the entanglement entropy only depends on the coefficient $h_{3,0}(t)$. We start by expanding \eqref{equation_two}. This gives us
	\begin{equation}
	0 =\mbox{ } (8\dot{h}_{2,0} + \phi_{0,0}\dot{\phi}_{0,0})z + (12\dot{h}_{3,0} + 2\phi_{2,0}\dot{\phi}_{0,0} + \phi_{0,0}\dot{\phi}_{2,0})z^{2} + O(z^{3}).
	\end{equation}
	To solve for $h_{3,0}(t)$, we first need to know $\phi_{2,0}(t)$. This can be easily found by expanding \eqref{equation_four}:
	\begin{equation}
	0 =\mbox{ }( \phi_{2,0} + \frac{\kappa}{4}\phi_{0,0}^{2}) z^{2} + O(z^{3})
	\end{equation}
	Solving the above two equations yields
	\begin{equation}
	h_{3,0}(t_{0}) =\mbox{ }\frac{\kappa}{36}\mbox{ }\phi_{0,0}^{3}(t_{0}). \label{res_a_one_d_five}
	\end{equation}
	
	\item {\bf $\alpha = 1$ and $d=6$}
	
	The entanglement entropy for this case is given by \eqref{a_one_d_six}. Here we need to solve for the combination $h_{2,0}^{2}(t) + 2h_{2,2}(t) + 2h_{4,0}(t)$. As always, we start by expanding \eqref{equation_two}. We get
	\begin{align}
	0 =\mbox{ }& (10 \dot{h}_{2,0} + \phi_{0,0}\dot{\phi}_{0,0})z + (15 \dot{h}_{3,0} + 2 \phi_{2,0}\dot{\phi}_{0,0} + \phi_{0,0}\dot{\phi}_{2,0})z^{2} + \Big( 20(\dot{h}_{2,2} + \dot{h}_{4,0}) \nonumber\\-& 5(f_{2,0}+3h_{3,0})\dot{h}_{2,0} + 3\dot{\phi}_{0,0}(\phi_{0,2}+\phi_{3,0}) + \phi_{0,0}(\dot{\phi}_{0,2}+\dot{\phi}_{3,0}) +2 \phi_{2,0}\dot{\phi}_{2,0}  \Big)z^{3} + O(z^{4}). \label{int_a_one_d_six}
	\end{align}
	Integrating the leading order term of this equation yields $h_{2,0}(t) = -\frac{1}{20}\phi_{0,0}^{2}(t)$. Furthermore, we can use the coefficient of the $z^{3}$ term to solve for the sum $h_{2,2}(t)+h_{3,0}(t)$. However, this would require us to solve for $f_{2,0}$, $\phi_{2,0}$, $\phi_{2,2}$ and $\phi_{3,0}$. All these can be found from the expansion of \eqref{equation_three} and \eqref{equation_four}. Expanding \eqref{equation_three} yields
	\begin{equation}
	0 =\mbox{ }(10 f_{2,0} +10 h_{2,0} + \phi_{0,0}^{2}) + O(z)
	\end{equation}
	which we solve to get $f_{2,0}(t) = -\frac{1}{20}\phi_{0,0}^{2}(t)$. Similarly, expanding the equation \eqref{equation_four} yields
	\begin{equation}
	0 =\mbox{ }(18\phi_{2,0} + 3\kappa \phi_{0,0}^{2})z^{2} + \Big( 24 (\phi_{0,2}+\phi_{3,0}) + 6\ddot{\phi}_{0,0} + \phi_{0,0}(-30h_{2,0}-6f_{2,0} + \omega \phi_{0,0}^{2} + 6\kappa \phi_{2,0})\Big)z^{3} + O(z^{4}).
	\end{equation}
	After using this equation to solve for $\phi_{2,0}$ and for $\phi_{0,2}+\phi_{3,0}$, we can solve for $h_{2,0}+h_{3,0}$ using \eqref{int_a_one_d_six}. Combining all the results, we get
	\begin{align}
	h_{2,0}^{2}(t) + 2h_{2,2}(t) + 2h_{4,0}(t) =\mbox{ } \frac{(117-65\kappa^{2} + 45\omega)}{7200}\mbox{ }\phi_{0,0}^{4}(t) + \frac{1}{80}\mbox{ }\partial_{t}^{2}\Big(\phi_{0,0}^{2}(t)\Big) .\label{res_a_one_d_six}
	\end{align}
	
	\end{itemize}


\section{Useful integral}
\label{app-integrations}

In this appendix we aim at simplifying the following integral 
\be\label{eq-N}
N_{s}(t) \equiv \int_{-t}^{0}dt' \int d\x' \int d\x_{\|}\int_{0}^{\infty}dx_{\perp}\mbox{ }x_{\perp} \lambda(t'+t) \bra{0}  T_{00}(0,x_{\perp},\x_{\|}){\phi}^{2}(t',\x')\ket{0}.
\ee
From \eqref{eq-Tphi2} and \eqref{eq-phiphi} we have
\be
\bra{0}  T_{00}(x) {\phi}^{2}(x')\ket{0}= \int \frac{d\p_1d\p_2}{(2\pi)^{2d-2}} \frac{-E_1E_2 - \p_1 \cdot \p_2 + m_0^2 + 2\xi (\p_1 + \p_2)^2}{4E_1E_2}e^{-i(p_1+p_2)\cdot(x-x')}.
\ee
Introducing $\mathbf{q} = \p_1 + \p_2$, $\Delta t = t-t'$, $\Delta x_\perp = x_\perp - x_\perp'$ and integrating \eqref{eq-N} with respect to $\x_\|$ and $\x_\|'$, yields
\begin{align}
N_{s}(t)  &= \mathcal{A}\int_{0}^{t}dt'\int dx_\perp'\int_{0}^{\infty}dx_{\perp}\mbox{ }x_{\perp} \lambda(t')\nonumber\\
&\times \int \frac{d\p_1dq_{\perp}}{(2\pi)^{d}} \left(\frac{-1 + p_{1\perp}^2  /E_1^2 + 4\xi }{8E_1^2}q_\perp^2 + O(q_\perp^3)\right)e^{-i(E_1+E_2)\Delta t + i q_\perp \Delta x_\perp}.
\end{align}

It is important that we perform the integrals over $x_\perp$ and $x_\perp'$ in the proper order so as to avoid an ambiguous ``$0\times \infty$" result. First we integrate over $x_\perp$, using the identity\footnote{We ignore $\delta'(q_\perp)$ in what follows, since it is multiplied by $q_\perp^n$ with $n\geq2$. If, however, one keeps these terms, then eventually they result in the ill defined distributions of the form $q^n\delta'(q_\perp)\delta(q_\perp)$ which vanish within, \eg dimensional regularization.}
\be\label{ind_pvalue}
\int_{0}^{\infty}dx_{\perp}\mbox{ }x_{\perp}e^{- i q_\perp x_\perp} = {\rm P.V.}\frac{-1}{q_\perp^2} + i\pi \delta'(q_\perp),
\ee
and then over $x_\perp'$ which results in a $2\pi \delta(q_\perp)$. The result is
\be\label{eq-answer}
N_{s}(t)  = \mathcal{A}\int_{0}^{t}dt' \lambda(t')\int \frac{d\p}{(2\pi)^{d-1}} \left(\frac{1 - p_{\perp}^2  /E^2 - 4\xi }{8E^2}\right)e^{2iE(t'-t)}.
\ee

\bibliographystyle{utcaps}
\bibliography{ent}

\providecommand{\href}[2]{#2}\begingroup\raggedright\begin{thebibliography}{10}

\bibitem{CC-particle}
P.~Calabrese and J.~L. Cardy, ``{Evolution of entanglement entropy in
  one-dimensional systems},''
  \href{http://dx.doi.org/10.1088/1742-5468/2005/04/P04010}{{\em J.Stat.Mech.}
  {\bf 0504} (2005)  P04010},
\href{http://arxiv.org/abs/cond-mat/0503393}{{\tt arXiv:cond-mat/0503393
  [cond-mat]}}.

\bibitem{Balasubramania_one}
V.~Balasubramanian, A.~Bernamonti, J.~de~Boer, N.~Copland, B.~Craps,
  E.~Keski-Vakkuri, B.~Muller, A.~Schafer, M.~Shigemori, and W.~Staessens,
  ``{Thermalization of Strongly Coupled Field Theories},''
  \href{http://dx.doi.org/10.1103/PhysRevLett.106.191601}{{\em Phys. Rev.
  Lett.} {\bf 106} (2011)  191601},
\href{http://arxiv.org/abs/1012.4753}{{\tt arXiv:1012.4753 [hep-th]}}.

\bibitem{Balasubramanian_two}
V.~Balasubramanian, A.~Bernamonti, J.~de~Boer, N.~Copland, B.~Craps,
  E.~Keski-Vakkuri, B.~Muller, A.~Schafer, M.~Shigemori, and W.~Staessens,
  ``{Holographic Thermalization},''
  \href{http://dx.doi.org/10.1103/PhysRevD.84.026010}{{\em Phys. Rev.} {\bf
  D84} (2011)  026010},
\href{http://arxiv.org/abs/1103.2683}{{\tt arXiv:1103.2683 [hep-th]}}.

\bibitem{Maldacena-Hartman}
T.~Hartman and J.~Maldacena, ``{Time Evolution of Entanglement Entropy from
  Black Hole Interiors},''
  \href{http://dx.doi.org/10.1007/JHEP05(2013)014}{{\em JHEP} {\bf 1305} (2013)
   014},
\href{http://arxiv.org/abs/1303.1080}{{\tt arXiv:1303.1080 [hep-th]}}.

\bibitem{HL-one}
H.~Liu and S.~J. Suh, ``{Entanglement Tsunami: Universal Scaling in Holographic
  Thermalization},''
  \href{http://dx.doi.org/10.1103/PhysRevLett.112.011601}{{\em Phys.Rev.Lett.}
  {\bf 112} (2014)  011601},
\href{http://arxiv.org/abs/1305.7244}{{\tt arXiv:1305.7244 [hep-th]}}.

\bibitem{HL-two}
H.~Liu and S.~J. Suh, ``{Entanglement growth during thermalization in
  holographic systems},''
  \href{http://dx.doi.org/10.1103/PhysRevD.89.066012}{{\em Phys.Rev.} {\bf D89}
  (2014) no.~6, 066012},
\href{http://arxiv.org/abs/1311.1200}{{\tt arXiv:1311.1200 [hep-th]}}.

\bibitem{tsunami}
S.~Leichenauer and M.~Moosa, ``{Entanglement Tsunami in (1+1)-Dimensions},''
  \href{http://dx.doi.org/10.1103/PhysRevD.92.126004}{{\em Phys. Rev.} {\bf
  D92} (2015)  126004},
\href{http://arxiv.org/abs/1505.04225}{{\tt arXiv:1505.04225 [hep-th]}}.

\bibitem{Asplund}
C.~T. Asplund, A.~Bernamonti, F.~Galli, and T.~Hartman, ``{Entanglement
  Scrambling in 2d Conformal Field Theory},''
  \href{http://dx.doi.org/10.1007/JHEP09(2015)110}{{\em JHEP} {\bf 09} (2015)
  110},
\href{http://arxiv.org/abs/1506.03772}{{\tt arXiv:1506.03772 [hep-th]}}.

\bibitem{CLM}
H.~Casini, H.~Liu, and M.~Mezei, ``{Spread of entanglement and causality},''
\href{http://arxiv.org/abs/1509.05044}{{\tt arXiv:1509.05044 [hep-th]}}.

\bibitem{Rangamani}
M.~Rangamani, M.~Rozali, and A.~Vincart-Emard, ``{Dynamics of Holographic
  Entanglement Entropy Following a Local Quench},''
\href{http://arxiv.org/abs/1512.03478}{{\tt arXiv:1512.03478 [hep-th]}}.

\bibitem{Bomb86}
L.~Bombelli, R.~K. Koul, J.~Lee, and R.~D. Sorkin, ``{A Quantum Source of
  Entropy for Black Holes},''
\href{http://dx.doi.org/10.1103/PhysRevD.34.373}{{\em Phys.Rev.} {\bf D34}
  (1986)  373--383}.

\bibitem{Sred93}
M.~Srednicki, ``{Entropy and area},''
  \href{http://dx.doi.org/10.1103/PhysRevLett.71.666}{{\em Phys.Rev.Lett.} {\bf
  71} (1993)  666--669},
\href{http://arxiv.org/abs/hep-th/9303048}{{\tt arXiv:hep-th/9303048
  [hep-th]}}.

\bibitem{misha_myers_hung}
L.-Y. Hung, R.~C. Myers, and M.~Smolkin, ``{Some Calculable Contributions to
  Holographic Entanglement Entropy},''
  \href{http://dx.doi.org/10.1007/JHEP08(2011)039}{{\em JHEP} {\bf 08} (2011)
  039},
\href{http://arxiv.org/abs/1105.6055}{{\tt arXiv:1105.6055 [hep-th]}}.

\bibitem{solo}
S.~N. Solodukhin, ``{Entanglement entropy, conformal invariance and extrinsic
  geometry},'' \href{http://dx.doi.org/10.1016/j.physletb.2008.05.071}{{\em
  Phys.Lett.} {\bf B665} (2008)  305--309},
\href{http://arxiv.org/abs/0802.3117}{{\tt arXiv:0802.3117 [hep-th]}}.

\bibitem{Hung:2011xb}
L.-Y. Hung, R.~C. Myers, and M.~Smolkin, ``{On Holographic Entanglement Entropy
  and Higher Curvature Gravity},''
  \href{http://dx.doi.org/10.1007/JHEP04(2011)025}{{\em JHEP} {\bf 1104} (2011)
   025},
\href{http://arxiv.org/abs/1101.5813}{{\tt arXiv:1101.5813 [hep-th]}}.

\bibitem{Fur13}
D.~V. Fursaev, A.~Patrushev, and S.~N. Solodukhin, ``{Distributional Geometry
  of Squashed Cones},''
  \href{http://dx.doi.org/10.1103/PhysRevD.88.044054}{{\em Phys.Rev.} {\bf D88}
  (2013) no.~4, 044054},
\href{http://arxiv.org/abs/1306.4000}{{\tt arXiv:1306.4000 [hep-th]}}.

\bibitem{Dong:2013qoa}
X.~Dong, ``{Holographic Entanglement Entropy for General Higher Derivative
  Gravity},'' \href{http://dx.doi.org/10.1007/JHEP01(2014)044}{{\em JHEP} {\bf
  1401} (2014)  044},
\href{http://arxiv.org/abs/1310.5713}{{\tt arXiv:1310.5713 [hep-th]}}.

\bibitem{Camps:2013zua}
J.~Camps, ``{Generalized entropy and higher derivative Gravity},''
  \href{http://dx.doi.org/10.1007/JHEP03(2014)070}{{\em JHEP} {\bf 03} (2014)
  070},
\href{http://arxiv.org/abs/1310.6659}{{\tt arXiv:1310.6659 [hep-th]}}.

\bibitem{Allais:2014ata}
A.~Allais and M.~Mezei, ``{Some results on the shape dependence of entanglement
  and R\'enyi entropies},''
\href{http://arxiv.org/abs/1407.7249}{{\tt arXiv:1407.7249 [hep-th]}}.

\bibitem{Mezei:2014zla}
M.~Mezei, ``{Entanglement entropy across a deformed sphere},''
\href{http://arxiv.org/abs/1411.7011}{{\tt arXiv:1411.7011 [hep-th]}}.

\bibitem{Faulkner:2015csl}
T.~Faulkner, R.~G. Leigh, and O.~Parrikar, ``{Shape Dependence of Entanglement
  Entropy in Conformal Field Theories},''
\href{http://arxiv.org/abs/1511.05179}{{\tt arXiv:1511.05179 [hep-th]}}.

\bibitem{Bianchi:2015liz}
L.~Bianchi, M.~Meineri, R.~C. Myers, and M.~Smolkin, ``{Renyi Entropy and
  Conformal Defects},''
\href{http://arxiv.org/abs/1511.06713}{{\tt arXiv:1511.06713 [hep-th]}}.

\bibitem{RS4}
V.~Rosenhaus and M.~Smolkin, ``{Entanglement Entropy for Relevant and Geometric
  Perturbations},'' \href{http://dx.doi.org/10.1007/JHEP02(2015)015}{{\em JHEP}
  {\bf 02} (2015)  015},
\href{http://arxiv.org/abs/1410.6530}{{\tt arXiv:1410.6530 [hep-th]}}.

\bibitem{Faulkner:2014jva}
T.~Faulkner, ``{Bulk Emergence and the RG Flow of Entanglement Entropy},''
  \href{http://dx.doi.org/10.1007/JHEP05(2015)033}{{\em JHEP} {\bf 05} (2015)
  033},
\href{http://arxiv.org/abs/1412.5648}{{\tt arXiv:1412.5648 [hep-th]}}.

\bibitem{RS1}
V.~Rosenhaus and M.~Smolkin, ``{Entanglement Entropy: A Perturbative
  Calculation},''
\href{http://arxiv.org/abs/1403.3733}{{\tt arXiv:1403.3733 [hep-th]}}.

\bibitem{RT}
S.~Ryu and T.~Takayanagi, ``{Holographic derivation of entanglement entropy
  from AdS/CFT},'' \href{http://dx.doi.org/10.1103/PhysRevLett.96.181602}{{\em
  Phys.Rev.Lett.} {\bf 96} (2006)  181602},
\href{http://arxiv.org/abs/hep-th/0603001}{{\tt arXiv:hep-th/0603001
  [hep-th]}}.

\bibitem{HRT}
V.~E. Hubeny, M.~Rangamani, and T.~Takayanagi, ``{A Covariant holographic
  entanglement entropy proposal},''
  \href{http://dx.doi.org/10.1088/1126-6708/2007/07/062}{{\em JHEP} {\bf 0707}
  (2007)  062},
\href{http://arxiv.org/abs/0705.0016}{{\tt arXiv:0705.0016 [hep-th]}}.

\bibitem{adscft_one}
S.~S. Gubser, I.~R. Klebanov, and A.~M. Polyakov, ``{Gauge theory correlators
  from noncritical string theory},''
  \href{http://dx.doi.org/10.1016/S0370-2693(98)00377-3}{{\em Phys. Lett.} {\bf
  B428} (1998)  105--114},
\href{http://arxiv.org/abs/hep-th/9802109}{{\tt arXiv:hep-th/9802109
  [hep-th]}}.

\bibitem{adscft_two}
E.~Witten, ``{Anti-de Sitter space and holography},'' {\em Adv. Theor. Math.
  Phys.} {\bf 2} (1998)  253--291,
\href{http://arxiv.org/abs/hep-th/9802150}{{\tt arXiv:hep-th/9802150
  [hep-th]}}.

\bibitem{Klebanov_Witten}
I.~R. Klebanov and E.~Witten, ``{AdS / CFT correspondence and symmetry
  breaking},'' \href{http://dx.doi.org/10.1016/S0550-3213(99)00387-9}{{\em
  Nucl. Phys.} {\bf B556} (1999)  89--114},
\href{http://arxiv.org/abs/hep-th/9905104}{{\tt arXiv:hep-th/9905104
  [hep-th]}}.

\bibitem{Witten_Graham}
C.~R. Graham and E.~Witten, ``{Conformal anomaly of submanifold observables in
  AdS / CFT correspondence},''
  \href{http://dx.doi.org/10.1016/S0550-3213(99)00055-3}{{\em Nucl. Phys.} {\bf
  B546} (1999)  52--64},
\href{http://arxiv.org/abs/hep-th/9901021}{{\tt arXiv:hep-th/9901021
  [hep-th]}}.

\bibitem{Cappelli:1990yc}
A.~Cappelli, D.~Friedan, and J.~I. Latorre, ``{C theorem and spectral
  representation},''
\href{http://dx.doi.org/10.1016/0550-3213(91)90102-4}{{\em Nucl.Phys.} {\bf
  B352} (1991)  616--670}.

\bibitem{Bis75}
J.~J. Bisognano and E.~H. Wichmann, ``{On the duality condition for a Hermitian
  scalar field},'' {\em Journal of Mathematical Physics} {\bf 16} (1975) no.~4,
  .

\bibitem{Bis76}
J.~J. Bisognano and E.~H. Wichmann, ``{On the duality condition for quantum
  fields},'' {\em Journal of Mathematical Physics} {\bf 17} (1976) no.~3, .

\bibitem{Kabat:1994vj}
D.~N. Kabat and M.~Strassler, ``{A Comment on entropy and area},''
  \href{http://dx.doi.org/10.1016/0370-2693(94)90515-0}{{\em Phys.Lett.} {\bf
  B329} (1994)  46--52},
\href{http://arxiv.org/abs/hep-th/9401125}{{\tt arXiv:hep-th/9401125
  [hep-th]}}.

\bibitem{Crisp07}
L.~C. Crispino, A.~Higuchi, and G.~E. Matsas, ``{The Unruh effect and its
  applications},'' \href{http://dx.doi.org/10.1103/RevModPhys.80.787}{{\em
  Rev.Mod.Phys.} {\bf 80} (2008)  787--838},
\href{http://arxiv.org/abs/0710.5373}{{\tt arXiv:0710.5373 [gr-qc]}}.

\bibitem{Hertzberg:2010uv}
M.~P. Hertzberg and F.~Wilczek, ``{Some Calculable Contributions to
  Entanglement Entropy},''
  \href{http://dx.doi.org/10.1103/PhysRevLett.106.050404}{{\em Phys.Rev.Lett.}
  {\bf 106} (2011)  050404},
\href{http://arxiv.org/abs/1007.0993}{{\tt arXiv:1007.0993 [hep-th]}}.

\bibitem{Huerta:2011qi}
M.~Huerta, ``{Numerical Determination of the Entanglement Entropy for Free
  Fields in the Cylinder},''
  \href{http://dx.doi.org/10.1016/j.physletb.2012.03.044}{{\em Phys.Lett.} {\bf
  B710} (2012)  691--696},
\href{http://arxiv.org/abs/1112.1277}{{\tt arXiv:1112.1277 [hep-th]}}.

\bibitem{Lewkowycz:2012qr}
A.~Lewkowycz, R.~C. Myers, and M.~Smolkin, ``{Observations on entanglement
  entropy in massive QFT's},''
  \href{http://dx.doi.org/10.1007/JHEP04(2013)017}{{\em JHEP} {\bf 1304} (2013)
   017},
\href{http://arxiv.org/abs/1210.6858}{{\tt arXiv:1210.6858 [hep-th]}}.

\bibitem{Casini:2014yca}
H.~Casini, F.~D. Mazzitelli, and E.~Test�, ``{Area terms in entanglement
  entropy},'' \href{http://dx.doi.org/10.1103/PhysRevD.91.104035}{{\em Phys.
  Rev.} {\bf D91} (2015) no.~10, 104035},
\href{http://arxiv.org/abs/1412.6522}{{\tt arXiv:1412.6522 [hep-th]}}.

\bibitem{Ben-Ami:2015zsa}
O.~Ben-Ami, D.~Carmi, and M.~Smolkin, ``{Renormalization group flow of
  entanglement entropy on spheres},''
  \href{http://dx.doi.org/10.1007/JHEP08(2015)048}{{\em JHEP} {\bf 08} (2015)
  048},
\href{http://arxiv.org/abs/1504.00913}{{\tt arXiv:1504.00913 [hep-th]}}.

\bibitem{Akers:2015bgh}
C.~Akers, O.~Ben-Ami, V.~Rosenhaus, M.~Smolkin, and S.~Yankielowicz,
  ``{Entanglement and RG in the $O(N)$ vector model},''
\href{http://arxiv.org/abs/1512.00791}{{\tt arXiv:1512.00791 [hep-th]}}.

\bibitem{Liu:2012eea}
H.~Liu and M.~Mezei, ``{A Refinement of entanglement entropy and the number of
  degrees of freedom},'' \href{http://dx.doi.org/10.1007/JHEP04(2013)162}{{\em
  JHEP} {\bf 1304} (2013)  162},
\href{http://arxiv.org/abs/1202.2070}{{\tt arXiv:1202.2070 [hep-th]}}.

\bibitem{Osborn:1993cr}
H.~Osborn and A.~Petkou, ``{Implications of conformal invariance in field
  theories for general dimensions},''
  \href{http://dx.doi.org/10.1006/aphy.1994.1045}{{\em Annals Phys.} {\bf 231}
  (1994)  311--362},
\href{http://arxiv.org/abs/hep-th/9307010}{{\tt arXiv:hep-th/9307010
  [hep-th]}}.

\bibitem{Erdmenger:1996yc}
J.~Erdmenger and H.~Osborn, ``{Conserved currents and the energy momentum
  tensor in conformally invariant theories for general dimensions},''
  \href{http://dx.doi.org/10.1016/S0550-3213(96)00545-7}{{\em Nucl.Phys.} {\bf
  B483} (1997)  431--474},
\href{http://arxiv.org/abs/hep-th/9605009}{{\tt arXiv:hep-th/9605009
  [hep-th]}}.

\bibitem{Smolkin:2014hba}
M.~Smolkin and S.~N. Solodukhin, ``{Correlation functions on conical
  defects},'' \href{http://dx.doi.org/10.1103/PhysRevD.91.044008}{{\em Phys.
  Rev.} {\bf D91} (2015) no.~4, 044008},
\href{http://arxiv.org/abs/1406.2512}{{\tt arXiv:1406.2512 [hep-th]}}.

\bibitem{RS3}
V.~Rosenhaus and M.~Smolkin, ``{Entanglement entropy, planar surfaces, and
  spectral functions},'' \href{http://dx.doi.org/10.1007/JHEP09(2014)119}{{\em
  JHEP} {\bf 09} (2014)  119},
\href{http://arxiv.org/abs/1407.2891}{{\tt arXiv:1407.2891 [hep-th]}}.

\bibitem{Lewkowycz:2013nqa}
A.~Lewkowycz and J.~Maldacena, ``{Generalized gravitational entropy},''
  \href{http://dx.doi.org/10.1007/JHEP08(2013)090}{{\em JHEP} {\bf 1308} (2013)
   090},
\href{http://arxiv.org/abs/1304.4926}{{\tt arXiv:1304.4926 [hep-th]}}.

\bibitem{Fursaev:2006ih}
D.~V. Fursaev, ``{Proof of the holographic formula for entanglement entropy},''
  \href{http://dx.doi.org/10.1088/1126-6708/2006/09/018}{{\em JHEP} {\bf 09}
  (2006)  018},
\href{http://arxiv.org/abs/hep-th/0606184}{{\tt arXiv:hep-th/0606184
  [hep-th]}}.

\bibitem{Headrick:2010zt}
M.~Headrick, ``{Entanglement Renyi entropies in holographic theories},''
  \href{http://dx.doi.org/10.1103/PhysRevD.82.126010}{{\em Phys. Rev.} {\bf
  D82} (2010)  126010},
\href{http://arxiv.org/abs/1006.0047}{{\tt arXiv:1006.0047 [hep-th]}}.

\bibitem{Dong:2016hjy}
X.~Dong, A.~Lewkowycz, and M.~Rangamani, ``{Deriving covariant holographic
  entanglement},''
\href{http://arxiv.org/abs/1607.07506}{{\tt arXiv:1607.07506 [hep-th]}}.

\end{thebibliography}\endgroup

\end{document}